\documentstyle[11pt]{article}
\begin{document}
\begin{large}
\begin{titlepage}

\vspace{0.2cm}

\title{One-loop quark and squark corrections to the lightest chargino pair
production in photon-photon collisions
\footnote{The project supported by National Natural Science
          Foundation of China}}
\author{{ Zhou Mian-Lai$^{b}$, Ma Wen-Gan$^{a,b,c}$, Han Liang$^{b}$,
          Jiang Yi$^{b}$ and Zhou Hong$^{b}$}\\
{\small $^{a}$CCAST (World Laboratory), P.O.Box 8730, Beijing 100080,
China.}\\
{\small $^{b}$Department of Modern Physics, University of Science
        and Technology}\\
{\small of China (USTC), Hefei, Anhui 230027, China.}\\
{\small $^{c}$Institute of Theoretical Physics, Academia Sinica,} \\
{\small P.O.Box 2735, Beijing 100080, China.} }
\date{}
\maketitle

\vskip 12mm

\begin{center}\begin{minipage}{5in}

\vskip 5mm
\begin{center} {\bf Abstract}\end{center}
\baselineskip 0.3in
{We present the detailed analytical and numerical
investigations of the one-loop radiative corrections of all quarks
and their supersymmetric partners in the MSSM to the lightest chargino pair
production via $\gamma \gamma$ fusion at the future NLC. The numerical
results show that the cross sections of the subprocess and the parent
process are typically enhanced by several percent compared to the results
at the lowest order, and these corrections are mainly contributed by the
virtual quarks and squarks of the third generation. Furthermore, we studied
the effects induced by the CP violating complex phases stemming from squark
and chargino mass matrices in the MSSM at one-loop level. We find that the
radiative corrections are related to all the three CP violating complex phases
$\phi_{t,b}$ and $\phi_{\mu}$. We conclude that the precise measurement of
the cross section and the experimental determination of the parameters
$\phi_{\mu}$ and $\phi_{t,b}$ are crucial in searching for SUSY signals. }\\

\vskip 5mm
{PACS number(s): 14.80.Ly, 12.15.Lk, 12.60.Jv}
\end{minipage}
\end{center}
\end{titlepage}

\baselineskip=0.36in

\eject
\rm
\baselineskip=0.36in

\begin{flushleft} {\bf 1. Introduction} \end{flushleft}
\par
  The Standard Model(SM) \cite{glash} \cite{higgs} can explain almost all
the currently available experimental data pertaining to the strong, weak
and electromagnetic interaction phenomena perfectly. The discovery of the
top quark in 1995 by the CDF and D0\cite{CDFD0} experiments at the Fermilab
Tevatron once again confirmed the standard model(SM). Only the elementary
Higgs boson, which is required strictly by the Standard Model for spontaneous
symmetry breaking, remains to be found. Therefore, the SM is a successful
theory of strong and electroweak interactions up to the present accessible
energies. At present, the supersymmetric extended model(SUSY)\cite{haber}
\cite{Gunion} is widely considered as theoretically the most appealing
extension of the SM. Apart from describing the experimental data well,
as the SM does, the supersymmetric theory is able to solve various
theoretical problems, such as the fact that the SUSY may provide an
elegant way to construct the huge hierarchy between the electroweak
symmetry-breaking scale and the grand unification scales.
\par
As we know, searching for sparticles directly at present and future
colliders is one of the promising tasks\cite{Han0}, and the accurate
measurements of the sparticle production process will give us significant
information about the MSSM. Like other sparticles, charginos can also be
produced in $e^{+} e^{-}$, $\gamma \gamma$ and hadron collisions. To detect
the existence of charginos, $e^{+}e^{-}$ and $\gamma \gamma$ collisions have
an advantage over hardron collisions due to its cleaner background.
So far there is no experimental evidence for charginos at LEP2. They only
set lower bounds on the lightest chargino mass $m_{\tilde{\chi}_{1}^{\pm}}$.
Recent experimental reports present that the mass of the lightest chargino may
be larger than $85~GeV$\cite{L3} \cite{ALEPH}\cite{OPAL}, and this bound depends
mainly on the sneutrino mass and the mass difference between the chargino
and the lightest SUSY particles(LSP) in theory.
\par
  The Next-generation Linear Collider(NLC) operated in laser back-scattering
photon collision mode at a c.m.s. energy of $500 \sim 2000~GeV$ with the
luminosity of the order of $10^{33} cm^{-2}s^{-1}$ may be an ideal instrument
to look for evidences of Higgs bosons and other new particles beyond the SM.
The analysis for the production of chargino pair ($\tilde{W}^{+} \tilde{W}^{-}$)
via $\gamma \gamma$ collision at tree-level
has been given in Ref.\cite{Goto}, and the production rate can be greater than
that by direct $e^+e^-$ annihilation, as the later has a `s-channel
suppression' due to the virtual photon propagator when the chargino is heavy.
Therefore, $\gamma \gamma$ collision provides another way to produce chargino
pair which is worth investigating.
\par
We know that the precise measurements of chargino pair production rates and
chargino masses give the possibility of measuring some gaugino,
higgsino couplings and constraining the mass scale of squarks, which might
not be in direct reach in colliders. Therefore, studying the chargino pair
production in photon-photon fusion process only at the tree-level is far
from the high precision requirements. Among all the radiative electroweak
corrections within the MSSM for the process $\gamma\gamma \rightarrow
\tilde{\chi}_{1}^{+} \tilde{\chi}_{1}^{-}$, the one-loop contributions of
the gaugino-higgsino-sector are very important. It is because not only
the Yukawa couplings of these heavy quarks and squarks might enhance the
correction, but also there exist the gaugino and higgsino couplings.
In previous studies, the tree-level and complete one-loop calculations of
heavy quarks and squarks for the chargino pair production in
electron-positron collisions have been performed in references
\cite{Diaz}\cite{Choi}. In Ref.\cite{Singo} S. Kiyoura et.al. calculated
the full quark and squark one-loop corrections to $e^+e^- \rightarrow
\tilde{\chi}^{+}_{1} \tilde{\chi}^{-}_{1}$ and made comparisons with
the results in \cite{Diaz}. There exist some numerical differences
between their results\cite{Singo}, but they both concluded
that the correction may be observable in chargino pair production at
$e^+e^-$ colliders.
\par
In this work, we will investigate the full one-loop quark and
squark corrections to the lightest chargino pair($\tilde{\chi}_{1}^{+}
\tilde{\chi}_{1}^{-}$) production via $\gamma \gamma$ fusion in the NLC
within the MSSM. The paper is organized as follows: In section II we
introduce the squark-sector and chargino-sector of the MSSM. In Sec.III
we give the analytical results for the cross sections of subprocess
$\gamma \gamma \rightarrow \tilde{\chi}_{1}^{+} \tilde{\chi}_{1}^{-}$
at tree-level and the leading one-loop corrections involving virtual top,
stop, bottom and sbottom quarks. In Sec.IV, the numerical results for
subprocess and parent process are illustrated, along with discussions.
Finally, a short summary is presented. In the Appendix, some lengthy
expressions of the form factors which appear in the cross section
in Sec.III are listed.

\begin{flushleft} {\bf 2. The squark-sector and chargino-sector of the
              MSSM and relevant Feynman rules.} \end{flushleft}
\par
In the MSSM theory every quark has two scalar partners, the squarks
$\tilde{q}_L$ and $\tilde{q}_R$. If there is no left-right flavor mixing
in the squark-sector, the mass matrix of a scalar quark including
CP-odd phases takes the following form\cite{s9}:
$$
 -{\cal L}_{m}=\left( \begin{array}{ll}
        \tilde{q}^{\ast}_{L} &  \tilde{q}^{\ast}_{R}
              \end{array} \right)
       \left( \begin{array}{ll}  m^2_{\tilde{q}_{L}} &  a_{q} m_{q} \\
           a_{q}^{\ast} m_{q}  &  m^2_{\tilde{q}_{R}}
              \end{array}  \right)
       \left( \begin{array}{ll}  \tilde{q}_{L}  \\  \tilde{q}_{R}
              \end{array}  \right),~~~~~~~~(2.1)
$$
where $\tilde{q}_{L}$ and $\tilde{q}_{R}$ are the current eigenstates and
for the up-type scalar quarks, we have
$$m^2_{\tilde{q}_{L}}=\tilde{M}^2_{Q} + m^2_{q} +
      m_{Z}^2 (\frac{1}{2}- Q_q s_{W}^2) \cos{2 \beta},~~~~~(2.2) $$
$$m^2_{\tilde{q}_{R}}=\tilde{M}^2_{U} + m^2_{q} +
      Q_q m_{Z}^2 s_{W}^2 \cos{2 \beta},~~~~~(2.3)  $$
$$
a_{q}=|a_{q}|e^{-2i \phi_{q}}=\mu \cot{\beta} + A^{\ast}_{q} \tilde{M}.~~~~~(2.4)
$$
For the down-type scalar quarks,
$$m^2_{\tilde{q}_{L}}=\tilde{M}^2_{Q} + m^2_{q}
      - m_{Z}^2 (\frac{1}{2}+ Q_q s_{W}^2) \cos{2 \beta} ,~~~~~(2.5) $$
$$m^2_{\tilde{q}_{R}}=\tilde{M}^2_{D} + m^2_{q}
      + Q_q m_{Z}^2 s_{W}^2 \cos{2 \beta},~~~~~(2.6)  $$
$$
a_{q}=|a_{q}|e^{-2i \phi_{q}}=\mu \tan{\beta} + A^{\ast}_{q} \tilde{M},~~~~~(2.7)
$$
where $Q_{q}$($Q_{D}=-\frac{1}{3}$, $Q_{U}=\frac{2}{3}$) is the
charge of the scalar quark, $\tilde{M}^2_{Q}$, $\tilde{M}^2_{U}$ and
$\tilde{M}^2_{D}$ are the self-supersymmetry-breaking mass terms for
the left-handed and
right-handed scalar quarks, $s_W=\sin{\theta_W}$, $c_W=\sin{\theta_W}$.
We choose $\tilde{M}_Q = \tilde{M}_U = \tilde{M}_D = \tilde{M}$. $A_{q}
\cdot \tilde{M}$ is a trilinear scalar interaction parameter, and $\mu$
is the supersymmetric mass mixing term of the Higgs boson. The complex
value $a_{q}$ can introduce CP-violation. In general, $\tilde{q}_L$ and
$\tilde{q}_R$ are mixed and give the mass eigenstates $\tilde{q}_1$
and $\tilde{q}_2$(usually we assume $m_{\tilde{q}_1} < m_{\tilde{q}_2}$).
The mass eigenstates $\tilde{q}_1$ and $\tilde{q}_2$ are expressed in
terms of the current eigenstates $\tilde{q}_L$, $\tilde{q}_R$ and the
mixing angle $\theta_{q}$ with the CP-violating phase $\phi_{q}$. They
read
$$
\tilde{q}_1=\tilde{q}_L \cos{\theta_q} e^{ i\phi_{q}} -
            \tilde{q}_R \sin{\theta_q} e^{-i\phi_{q}},
~~~~~~~~~~~~~~
$$
$$
\tilde{q}_2= \tilde{q}_L \sin{\theta_q} e^{ i\phi_{q}} +
              \tilde{q}_R \cos{\theta_q} e^{-i\phi_{q}},
~~~~~~~~~(2.8)
$$
and
$$
\tan{2 \theta_q}=\frac{2 |a_{q}| m_{q}}
                      {m^2_{\tilde{q}_{L}}-m^2_{\tilde{q}_{R}} }.
~~~~~~~~~(2.9)
$$
\par
Then the masses of $\tilde{q}_1$ and $\tilde{q}_2$ are
$$
(m^2_{\tilde{q}_1},m^2_{\tilde{q}_2})=\frac{1}{2} \{
     m^2_{\tilde{q}_L} + m^2_{\tilde{q}_R} \mp  [ (m^2_{\tilde{q}_L} -
     m^2_{\tilde{q}_R})^2 + 4 |a_q|^2 m_q^2 ] ^{\frac{1}{2}} \}.~~~~~~(2.10)
$$
   The charginos $\tilde{\chi}_{i}^{+}~~(i=1,2)$ are four-component Dirac
fermions which arise due to the mixing of the SUSY partners of the charged
Higgs (charged Higgsinos $\tilde{H}_{1}^{-}$ and $\tilde{H}_{2}^{+}$) and
the $W$ gauge bosons(the winos $\tilde{W}^{\pm}$). The chargino mass term
in Lagrangian has the form
$$
 {\cal L}_{m}=-\frac{1}{2}(\psi^{+}~ \psi^{-}) \left(
      \begin{array}{ll}
        0  &  X^{T}  \\
        X  &  0
      \end{array}
      \right) \left(
      \begin{array}{l}
      \psi^{+}  \\
      \psi^{-}
      \end{array}
      \right).
\eqno{(2.11)}
$$
 where
$$
      X = \left(
      \begin{array}{ll}
         M_{SU(2)} & m_{W}\sqrt{2}\sin\beta  \\
         m_{W}\sqrt{2}\cos\beta &  |\mu|e^{i\phi_{\mu}}
      \end{array}
      \right),
\eqno{(2.12)}
$$
and we set $M_{SU(2)}$ to be real because its complex phase angle can be
rotated away by the field transformation and hence absorbed in $\phi_{\mu}$.
The two $2 \times 2$ unitary matrices U, V are defined to diagonalize the
matrix $X$, namely,
$$
      U^{\ast}XV^{\dag} = X_{D},
\eqno{(2.13)}
$$
where $X_{D}$ is a diagonal matrix with two non-negative entries $M_{\pm}$
standing for the chargino masses $m_{\tilde{\chi}^{+}_{1,2}}$ at
the tree level. The two diagonal elements of $X_{D}$ are worked out in
general case as
\begin{eqnarray*}
      M_{\pm}^{2} &=& \frac{1}{2}
        \left\{ M_{SU(2)}^{2}  + |\mu|^{2} + 2 m_{W}^{2} \pm
        \left[ (M_{SU(2)}^{2}-|\mu|^{2})^{2} + 4 m_{W}^{4} \cos^{2}2\beta +
        \right. \right. \\
&& \left. \left. 4 m_{W}^{2} (M_{SU(2)}^{2} +
         |\mu|^{2}+ 2 M_{SU(2)} |\mu| \sin2\beta \cos\phi_{\mu})
                     \right]^{1/2} \right\},~~~~~(2.14)
\end{eqnarray*}
Then the fundamental SUSY parameters $M_{SU(2)}$ and $|\mu|$ can be obtained
from the alternative expressions on the right-hand side of the following
equation, respectively.
$$
(M_{SU(2)}, |\mu|)=\frac{1}{2} \left(
   \sqrt{m^2_{\tilde{\chi}^{+}_{1}}+m^2_{\tilde{\chi}^{+}_{2}}-2 m_W^2+2 M_c}
   \pm \sqrt{m^2_{\tilde{\chi}^{+}_{1}}+m^2_{\tilde{\chi}^{+}_{2}}-
     2 m_W^2-2 M_c} \right),
~~~(2.15.1)
$$
where
$$
M_c= m_W^2 \cos{\phi_{\mu}} \sin{2 \beta}+\sqrt{m^2_{\tilde{\chi}^{+}_{1}} m^2_{\tilde{\chi}^{+}_{2}}-m_W^4
     \sin^2 2\beta \sin^2\phi_{\mu} }.~~~~~~(2.15.2)
$$
The diagonalizing matrices U and V generally have the forms:
$$
      U = \left(
      \begin{array}{ll}
         \cos\theta_{U}e^{i(\phi_{1}+\xi_{1})} &
         \sin\theta_{U}e^{i(\phi_{1}+\xi_{1}+\delta_{U})} \\
        -\sin\theta_{U}e^{i(\phi_{2}+\xi_{2}-\delta_{U})} &
         \cos\theta_{U}e^{i(\phi_{2}+\xi_{2})}
      \end{array}
      \right)
$$
$$
      V = \left(
      \begin{array}{ll}
         \cos\theta_{V}e^{i(\phi_{1}-\xi_{1})} &
         \sin\theta_{V}e^{i(\phi_{1}-\xi_{1}+\delta_{V})} \\
        -\sin\theta_{V}e^{i(\phi_{2}-\xi_{2}-\delta_{V})} &
         \cos\theta_{V}e^{i(\phi_{2}-\xi_{2})}
      \end{array}
      \right),
\eqno{(2.16)}
$$
      where the $\xi_{1}$ and $\xi_{2}$ can be any arbitrarily chosen phases.
It indicates that the matrices U and V satisfying Eq.(2.13) are not unique,
namely, some arbitrary phases can be introduced but they have no physical
effect. The explicit forms of the related constant angles and phases
depending on the input parameters are given as
$$
\tan{\theta_{U}}=\sqrt{\frac{M_{+}^{2}-M_{SU(2)}^{2}-2m_{W}^{2}\sin^{2}\beta}
            {M_{+}^{2}-|\mu|^{2}-2m_{W}^{2}\cos^{2}\beta}},
$$
$$
\tan{\theta_{V}}=\sqrt{\frac{M_{+}^{2}-M_{SU(2)}^{2}-2m_{W}^{2}\cos^{2}\beta}
            {M_{+}^{2}-|\mu|^{2}-2m_{W}^{2}\sin^{2}\beta}},
$$
$$
e^{i2\phi_{1}}=\frac{\cos\theta_{U}}{\cos\theta_{V}} \cdot
                     \frac{M_{+}^{2}+M_{SU(2)}|\mu|\tan\beta e^{i\phi_{\mu}}
                                   -2m_{W}^{2}\sin^{2}\beta}
                         {M_{+}(M_{SU(2)}+|\mu| \tan\beta e^{i\phi_{\mu}})},
$$
$$
e^{i2\phi_{2}}=\frac{\cos\theta_{V}}{\cos\theta_{U}} \cdot
                     \frac{M_{-}^{2}+M_{SU(2)}|\mu|\tan\beta e^{i\phi_{\mu}}
                                   -2m_{W}^{2}\sin^{2}\beta}
                         {M_{-}(M_{SU(2)}\tan\beta+|\mu| e^{-i\phi_{\mu}})},
$$
$$
      e^{i\delta_{U}}=\frac{M_{SU(2)}+|\mu| e^{i\phi_{\mu}}\tan\beta}
                           {|M_{SU(2)}+|\mu| e^{i\phi_{\mu}}\tan\beta|},
$$
$$
      e^{i\delta_{V}}=\frac{M_{SU(2)}\tan\beta+|\mu| e^{i\phi_{\mu}}}
                           {|M_{SU(2)}\tan\beta+|\mu| e^{i\phi_{\mu}}|},
\eqno{(2.17)}
$$
where $M_{\pm}$ can be evaluated from Eq.(2.14). In the MSSM, there are
many approaches to introduce CP-odd phases\cite{dim}. In our calculation,
only two kinds of CP-odd phases, respectively appearing in the squark mass
and chargino mass matrices, are involved. Although the detailed analyses of
the present upper bounds on electron and neutron electric dipole moments may
give constraints on CP-odd phase parameters indirectly\cite{Ibr}, yet these
constraints should be rather weak, since they depend strongly on the
assumptions to be applied. Recently S.Y. Choi et al discussed the impacts
of the CP-odd phase stemming from chargino mass matrix in the production of
the lightest chargino-pair in $e^+ e^-$ collisions at tree-level\cite{Achoi}.
In our work we are to investigate the effects from the CP-odd phases in squark
mass and chargino mass matrices in the process of the lightest chargino pair
production in $\gamma \gamma$ collisions at one-loop level. So we keep
all the relevant CP-odd complex phases and do not put any extra limitations
on CP-odd phases for the general discussion in our calculation.

  The Feynman rules for the couplings of $q-\tilde{q}^{\prime}_{L,R}-
\tilde{\chi}^{+}_1$ are presented in Ref.\cite{haber}\cite{Gunion}. Then
we can obtain the corresponding Feynman rules for such vertices in squark
mass eigenstate basis(see Fig.2). We denote the couplings in Fig.2 in
the forms of
$$
\bar{U}-\tilde{D}_{i}-\tilde{\chi}^{+}_{j}:~~
V_{U\tilde{D}_{i}\tilde{\chi}^{+}_{j}}^{(1)}P_L+
V_{U\tilde{D}_{i}\tilde{\chi}^{+}_{j}}^{(2)}P_R,
\eqno{(2.18.1)}
$$
$$
U-\bar{\tilde{D}}_{i}-\bar{\tilde{\chi}}^{+}_{j}:~~
-V_{U\tilde{D}_{i}\tilde{\chi}^{+}_{j}}^{(2)\ast}P_L-
V_{U\tilde{D}_{i}\tilde{\chi}^{+}_{j}}^{(1)\ast}P_R,
\eqno{(2.18.2)}
$$

$$
D-\bar{\tilde{U}}_{i}-\bar{\tilde{\chi}}^{+c}_{j}:~~
C^{-1}\left\{ V_{D\tilde{U}_{i}\tilde{\chi}^{+}_{j}}^{(1)}P_L+
V_{D\tilde{U}_{i}\tilde{\chi}^{+}_{j}}^{(2)}P_R \right\},
\eqno{(2.18.3)}
$$
$$
\bar{D}-\tilde{U}_{i}-\tilde{\chi}^{+c}_{j}:~~
\left\{ V_{D\tilde{U}_{i}\tilde{\chi}^{+}_{j}}^{(2)\ast}P_L+
V_{D\tilde{U}_{i}\tilde{\chi}^{+}_{j}}^{(1)\ast}P_R \right\} C,
\eqno{(2.18.4)}
$$
respectively. Here $(U,D)=(u,d),(c,s),(t,b)$ and C is the charge conjugation
matrix, which appears when there is a discontinuous flow of fermion number,
$P_{L,R}=\frac{1}{2} (1\mp\gamma_5)$ and
$$
V_{U\tilde{D}_{1}\tilde{\chi}^{+}_{j}}^{(1)}= \frac{i g m_U}
  {\sqrt{2}m_W \sin{\beta}} V_{j2}^{\ast} \cos{\theta_{D}} e^{-i\phi_D},
\eqno{(2.19.1)}
$$
$$
V_{U\tilde{D}_{1}\tilde{\chi}^{+}_{j}}^{(2)}= -i g (U_{j1} \cos{\theta_D}
   e^{-i\phi_D} + \frac{m_D}{\sqrt{2}m_W \cos{\beta}} U_{j2} \sin{\theta_D}
   e^{i\phi_D}),
\eqno{(2.19.2)}
$$
$$
V_{U\tilde{D}_{2}\tilde{\chi}^{+}_{j}}^{(1)}= \frac{i g m_U}
  {\sqrt{2}m_W \sin{\beta}} V_{j2}^{\ast} \sin{\theta_{D}} e^{-i\phi_D},
\eqno{(2.19.3)}
$$
$$
V_{U\tilde{D}_{2}\tilde{\chi}^{+}_{j}}^{(2)}= -i g (U_{j1} \sin{\theta_D}
   e^{-i\phi_D}- \frac{m_D}{\sqrt{2}m_W \cos{\beta}} U_{j2} \cos{\theta_D}
   e^{i\phi_D}),
\eqno{(2.19.4)}
$$
$$
V_{D\tilde{U}_{1}\tilde{\chi}^{+}_{j}}^{(1)}= ig (V_{j1}^{\ast} \cos{\theta_U}
   e^{i\phi_U} + \frac{m_U}{\sqrt{2}m_W \sin{\beta}} V_{j2}^{\ast}
   \sin{\theta_U} e^{-i\phi_D}),
\eqno{(2.20.1)}
$$
$$
V_{D\tilde{U}_{1}\tilde{\chi}^{+}_{j}}^{(2)}= \frac{-igm_D}{\sqrt{2}m_W \cos{\beta}}
   U_{j2} \cos{\theta_U} e^{i\phi_D},
\eqno{(2.20.2)}
$$
$$
V_{D\tilde{U}_{2}\tilde{\chi}^{+}_{j}}^{(1)}= ig (V_{j1}^{\ast} \sin{\theta_U}
   e^{i\phi_U} - \frac{m_U}{\sqrt{2}m_W \sin{\beta}} V_{j2}^{\ast}
   \cos{\theta_U} e^{-i\phi_D}),
\eqno{(2.20.3)}
$$
$$
V_{D\tilde{U}_{2}\tilde{\chi}^{+}_{j}}^{(2)}= \frac{igm_D}{\sqrt{2}m_W \cos{\beta}}
   U_{i2} \sin{\theta_U} e^{-i\phi_D},
\eqno{(2.20.4)}
$$
  For the Feynman rules of the Higgs-quark-quark, Higgs-squark-squark,
Higgs-chargino-chargino and Z$(\gamma)$-chargino-chargino, one can refer
to Ref.\cite{haber}\cite{Gunion}.
The couplings of $Higgs(B)-\tilde{\chi}^{+}_{k}-\tilde{\chi}^{+}_{k}$ are
\begin{eqnarray*}
V_{B\tilde{\chi}^{+}_{k}\tilde{\chi}^{+}_{k}}=
V_{B\tilde{\chi}^{+}_{k}\tilde{\chi}^{+}_{k}}^{s}+
V_{B\tilde{\chi}^{+}_{k}\tilde{\chi}^{+}_{k}}^{ps} \gamma_5~~
(B=h^0, H^0, A^0, G^0),~~~~~(2.21.1)
\end{eqnarray*}
where the notations defined above, which are involved in our calculation,
are explicitly expressed as below:
$$
V_{H^0\tilde{\chi}^{+}_{k}\tilde{\chi}^{+}_{k}}^{s} =
  \frac{-i g}{\sqrt{2}} \left[ \cos\alpha Re(V_{k,1} U_{k,2}) +
  \sin\alpha Re(V_{k,2} U_{k,1}) \right]
\eqno{(2.21.2)}
$$
$$
V_{h^0\tilde{\chi}^{+}_{k}\tilde{\chi}^{+}_{k}}^{s} =
   \frac{i g}{\sqrt{2}} \left[ \sin\alpha Re(V_{k,1} U_{k,2}) -
   \cos\alpha Re(V_{k,2} U_{k,1}) \right]
\eqno{(2.21.3)}
$$
$$
V_{A^0\tilde{\chi}^{+}_{k}\tilde{\chi}^{+}_{k}}^{ps} =
   \frac{g}{\sqrt{2}} \left[ \sin\beta Re(V_{k,1} U_{k,2}) +
   \cos\beta Re(V_{k,2} U_{k,1}) \right]
\eqno{(2.21.4)}
$$
$$
V_{G^0\tilde{\chi}^{+}_{k}\tilde{\chi}^{+}_{k}}^{ps} =
   \frac{-g}{\sqrt{2}} \left[ \cos\beta Re(V_{k,1} U_{k,2}) -
   \sin\beta Re(V_{k,2} U_{k,1}) \right]
\eqno{(2.21.5)}
$$
We define the following notations in Higgs-quark-quark and Higgs-squark-squark
couplings:
$$
H^0-U-U:~~
V_{H^{0}UU} = \frac{-i g m_U \sin{\alpha}}{2 m_W \sin{\beta}},~~~~
H^0-D-D:~~
V_{H^{0}DD} = \frac{-i g m_D \cos{\alpha}}{2 m_W \cos{\beta}},
\eqno{(2.22.1)}
$$
$$
h^0-U-U:~~
V_{h^{0}UU} = \frac{-i g m_U \cos{\alpha}}{2 m_W \sin{\beta}},~~
h^0-D-D:~~
V_{h^{0}DD} = \frac{i g m_D \sin{\alpha}}{2 m_W \cos{\delta}},
\eqno{(2.22.2)}
$$
$$
A^0-U-U:~~
V_{A^{0}UU}\gamma_5 = \frac{-g m_U \cot{\beta}}{2 m_W}\gamma_5,~~~~
A^0-D-D:~~
V_{A^{0}DD}\gamma_5 = \frac{-g m_D \tan{\beta}}{2 m_W}\gamma_5,
\eqno{(2.22.3)}
$$
$$
G^0-U-U:~~
V_{G^{0}UU}\gamma_5 = \frac{-g m_U}{2 m_W}\gamma_5,~~~~
G^0-D-D:~~
V_{G^{0}DD}\gamma_5 = \frac{g m_D}{2 m_W}\gamma_5.
\eqno{(2.22.4)}
$$
The couplings of $H^0(h^0)-\tilde{q}_{i}-\tilde{q}_{i}~~(i=1,2,q=u,d,c,s,t,b)$
are
\begin{eqnarray*}
V_{H^0\tilde{U}_{1}\tilde{U}_{1}} &=&
  \frac{-i g m_Z \cos{(\alpha+\beta)}}{\cos{\theta_W}} \left[
   (\frac{1}{2} - \frac{2}{3} \sin^2 \theta_W) \cos^2\theta_U +
    \frac{2}{3} \sin^2 \theta_W \sin^2\theta_U \right] \\
&& - \frac{i g m_U^2 \sin{\alpha}}{m_W \sin\beta}
   + \frac{i g m_U}{2 m_W \sin{\beta}} (A_U \sin{\alpha} + \mu \cos{\alpha})
     \sin{\theta_U} \cos{\theta_U} \cos{ 2 \phi_U},~~~~(2.23.1)
\end{eqnarray*}
\begin{eqnarray*}
V_{H^0\tilde{U}_{2}\tilde{U}_{2}} &=&
  \frac{-i g m_Z \cos(\alpha+\beta)}{\cos\theta_W} \left[
 (\frac{1}{2} - \frac{2}{3} \sin^2 \theta_W) \sin^2\theta_U +
  \frac{2}{3} \sin^2 \theta_W \cos^2\theta_U \right] \\
&& - \frac{i g m_U^2 \sin\alpha}{m_W \sin\beta}
   - \frac{i g m_U}{2 m_W \sin\beta} (A_U \sin\alpha + \mu \cos\alpha)
     \sin\theta_U \cos\theta_U \cos 2 \phi_U,~~~~(2.23.2)
\end{eqnarray*}
\begin{eqnarray*}
V_{H^0\tilde{D}_{1}\tilde{D}_{1}} &=&
  \frac{i g m_Z \cos(\alpha+\beta)}{\cos\theta_W} \left[
 (\frac{1}{2} - \frac{1}{3} \sin^2 \theta_W) \cos^2\theta_D +
  \frac{1}{3} \sin^2 \theta_W \sin^2\theta_D \right] \\
&& - \frac{i g m_D^2 \cos\alpha}{m_W \cos\beta}
   + \frac{i g m_D}{2 m_W \cos\beta} (A_D \cos\alpha + \mu \sin\alpha)
     \sin\theta_D \cos\theta_D \cos 2 \phi_D,~~~~(2.23.3)
\end{eqnarray*}
\begin{eqnarray*}
V_{H^0\tilde{D}_{2}\tilde{D}_{2}} &=&
  \frac{i g m_Z \cos(\alpha+\beta)}{\cos\theta_W} \left[
 (\frac{1}{2} - \frac{1}{3} \sin^2 \theta_W) \sin^2\theta_D +
  \frac{1}{3} \sin^2 \theta_W \cos^2\theta_D \right] \\
&& - \frac{i g m_D^2 \cos\alpha}{m_W \cos\beta}
   - \frac{i g m_D}{2 m_W \cos\beta} (A_D \cos\alpha + \mu \sin\alpha)
     \sin\theta_D \cos\theta_D \cos 2 \phi_D,~~~~(2.23.4)
\end{eqnarray*}
\begin{eqnarray*}
V_{h^0\tilde{U}_{1}\tilde{U}_{1}} &=&
  \frac{i g m_Z \sin(\alpha+\beta)}{\cos\theta_W} \left[
 (\frac{1}{2} - \frac{2}{3} \sin^2 \theta_W) \cos^2\theta_U +
  \frac{2}{3} \sin^2 \theta_W \sin^2\theta_U \right] \\
&& - \frac{i g m_U^2 \cos\alpha}{m_W \sin\beta}
   + \frac{i g m_U}{2 m_W \sin\beta} (A_U \cos\alpha - \mu \sin\alpha)
     \sin\theta_U \cos\theta_U \cos 2 \phi_U,~~~~(2.23.5)
\end{eqnarray*}
\begin{eqnarray*}
V_{h^0\tilde{U}_{2}\tilde{U}_{2}} &=&
  \frac{i g m_Z \sin(\alpha+\beta)}{\cos\theta_W} \left[
 (\frac{1}{2} - \frac{2}{3} \sin^2 \theta_W) \sin^2\theta_U +
  \frac{2}{3} \sin^2 \theta_W \cos^2\theta_U \right] \\
&& - \frac{i g m_U^2 \cos\alpha}{m_W \sin\beta}
   - \frac{i g m_U}{2 m_W \sin\beta} (A_U \cos\alpha - \mu \sin\alpha)
     \sin\theta_U \cos\theta_U \cos 2 \phi_U,~~~~(2.23.6)
\end{eqnarray*}
\begin{eqnarray*}
V_{h^0\tilde{D}_{1}\tilde{D}_{1}} &=&
 \frac{-i g m_Z \sin(\alpha+\beta)}{\cos\theta_W} \left[
(\frac{1}{2} - \frac{1}{3} \sin^2 \theta_W) \cos^2\theta_D +
 \frac{1}{3} \sin^2 \theta_W \sin^2\theta_D \right] \\
&& + \frac{i g m_D^2 \sin\alpha}{m_W \cos\beta}
   - \frac{i g m_D}{2 m_W \cos\beta} (A_D \sin\alpha - \mu \cos\alpha)
     \sin\theta_D \cos\theta_D \cos 2 \phi_D,~~~~(2.23.7)
\end{eqnarray*}
\begin{eqnarray*}
V_{h^0\tilde{D}_{2}\tilde{D}_{2}} &=&
  \frac{-i g m_Z \sin(\alpha+\beta)}{\cos\theta_W} \left[
 (\frac{1}{2} - \frac{1}{3} \sin^2 \theta_W) \sin^2\theta_D +
  \frac{1}{3} \sin^2 \theta_W \cos^2\theta_D \right] \\
&& + \frac{i g m_D^2 \sin\alpha}{m_W \cos\beta}
   + \frac{i g m_D}{2 m_W \cos\beta} (A_D \sin\alpha - \mu \cos\alpha)
     \sin\theta_D \cos\theta_D \cos 2 \phi_D,~~~~(2.23.8)
\end{eqnarray*}
respectively.

\par
\begin{flushleft} {\bf 3. The calculation of subprocess $\gamma \gamma
      \rightarrow \tilde{\chi}_1^{+} \tilde{\chi}_1^{-}$ in the MSSM.}
      \end{flushleft}
\par
Charginos are produced in the t- and u-channel with intermediate charginos
at the lowest order. The Feynman diagrams for the process at the tree level
are shown in Fig.1(a), where the u-channel tree-level is not drawn.
The relevant Feynman rules can be found in reference\cite{haber}.
The one-loop diagrams involving quarks and their supersymmetric partners
for the subprocess $\gamma \gamma \rightarrow \tilde{\chi}^{+}_1
\tilde{\chi}^{-}_1$, are drawn in Fig.1 ($b \sim g$), where we only give
the third generation quark and squark loop diagrams. All the one-loop
diagrams can be divided into five groups: (1) $\gamma \tilde{\chi}_1^{+}
\tilde{\chi}_1^{-}$ vertex correction diagrams shown as Fig.1(b). (2) box
diagrams as Fig.1(c). (3) quartic interaction diagrams as Fig.1(d). (4)
triangle diagrams as Fig.1(e). (5) self-energy correction diagram as
Fig.1(f). The chargino, $\gamma$ and $Z^{0}-\gamma$ mixing self-energy
diagrams shown as Fig.1(g). The total cross section including the leading
one-loop corrections in the frame of the MSSM should be
$$
\hat{\sigma} = \hat{\sigma}_{0} + \delta \hat{\sigma}^{1-loop},
$$
where $\delta \hat{\sigma}^{1-loop}$ represents the cross section with
all virtual quarks and their SUSY partners corrections. In
the calculation, we use the t'Hooft gauge and adopt the dimensional reduction
scheme(DR) \cite{Copper}, which is commonly used in the calculation of the
radiative corrections in frame of the MSSM as it preserves supersymmetry at
least at one-loop order, to control the ultraviolet divergences in the
virtual loop corrections. We choose the on-mass-shell scheme (OMS)\cite{se}
in doing renormalization.

\par
\begin{center} {\bf 3.1 The tree-level formulae and notations.}\end{center}

\par
  In this work, we denote the reaction of chargino pair production
via photon-photon collision as:
$$
\gamma (p_3, \mu) \gamma (p_4, \nu) \longrightarrow
            \tilde{\chi}_{1}^{+} (p_1) \tilde{\chi}_{1}^{-} (p_2).
\eqno{(3.1.1)}
$$
where $p_1$ and $p_2$ represent the momenta of the outgoing chargino pair and
$p_3$ and $p_4$ describe the momenta of the two incoming photons, respectively.
The Mandelstam variables $\hat{s}$, $\hat{t}$ and $\hat{u}$ are defined
as $\hat{s}=(p_{1}+p_{2})^2,~~\hat{t}=(p_1-p_3)^2,~~\hat{u}=(p_1-p_4)^2$.
The corresponding Lorentz invariant matrix element at the lowest order for the
reaction $\gamma \gamma \rightarrow \tilde{\chi}_{1}^{+} \tilde{\chi}_{1}^{-}$
is written as
$$
{\cal M}_{0}={\cal M}_{\hat{t}}+{\cal M}_{\hat{u}},
\eqno{(3.1.2)}
$$
where
$$
{\cal M}_{\hat{t}}=\left[ \bar{u}(p_3)(-i e \gamma_{\mu})
\frac{i}{\hat{\rlap/t}-m_{\tilde{\chi}_{1}^{+}}} (-i e \gamma_{\nu}) v(p_4)
\epsilon^{\mu}(p1) \epsilon^{\nu}(p2) \right],
\eqno{(3.1.3)}
$$
$$
{\cal M}_{\hat{u}}=\left[ \bar{u}(p_3)(-i e \gamma_{\nu})
\frac{i}{\hat{\rlap/u}-m_{\tilde{\chi}_{1}^{+}}} (-i e \gamma_{\mu}) v(p_4)
\epsilon^{\nu}(p2) \epsilon^{\mu}(p1) \right].
\eqno{(3.1.4)}
$$
The corresponding differential cross section is obtained by
$$
\frac{d \hat{\sigma}_{0}(\hat{t},\hat{s})} {d \hat{t}}=
   \frac{1}{16\pi^2\hat{s}}\bar{\sum_{spins}}|{\cal M}_{0}|^2,
\eqno{(3.1.5)}
$$
where the bar over the summation means to sum up the spins of final states
and average the spins of initial photons. After integration over $\hat{t}$,
the total Born cross section with unpolarized incoming photons is
given by\cite{Goto}
$$
\hat{\sigma}_{0}(\hat{s})=\frac{2\pi \alpha^2}{\hat{s}} \left[
           -2\hat{\beta}(2-\hat{\beta}^2) +(3-\hat{\beta}^4)
           \ln{ \frac{1+\hat{\beta}}{1-\hat{\beta}} }\right].
\eqno{(3.1.6)}
$$
where the kinematic factor is defined as
$$
\hat{\beta}=\sqrt{ 1-4 m_{\tilde{\chi}_{1}^{+}}^2/\hat{s} }.
\eqno{(3.1.7)}
$$

\par
\begin{center} {\bf 3.2 Self-energies.}
\end{center}

\par
  The self-energies of $\gamma$ and its mixing with $Z^0$ contributed by
the one-loop diagrams of virtual quarks and squarks shown in Fig.1(g.2)
read:
\begin{eqnarray*}
\Sigma_{T}^{AA}(p^2) &=& \frac{N_c e^2}{8 \pi^2}
   \sum_{q=(u,d)}^{(c,s),(t,b)} \left\{
   2 Q_q^2 (m_q^2 B_{0}^{q} - p^2 B_{1}^{q} - p^2 B_{21}^{q} +
   (2-\epsilon) B_{22}^{q}) \right. \\
&&\left. + \sum_{i=1,2} \left[ Q_q^2 (2 B_{22}^{\tilde{q}_{i}}
   - A_{0}(m_{\tilde{q}_{i}}) ) \right] \right\}, ~~~~~~~~~(3.2.1)
\end{eqnarray*}
\begin{eqnarray*}
\Sigma_{T}^{AZ}(p^2) &=&
  \frac{N_c e g}{16 \pi^2 c_{W}} \sum_{(U,D)=(u,d)}^{(c,s),(t,b)} \left\{
    Q_D (cos^2{\theta_D}+2 Q_D s_W^2) (A_{0}(m_{\tilde{D}_1})-
    (2-\epsilon) B_{22}^{\tilde{D}_1}) \right. \\
&+& \left. Q_D (sin^2{\theta_D}+2 Q_D s_W^2) (A_{0}(m_{\tilde{D}_2})-
    B_{22}^{\tilde{D}_2}) \right. \\
&-& \left.  Q_D (1+4 Q_D s_W^2) (m_D^2  B_{0}^{D} - p^2 B_{1}^{D}
    - p^2 B_{21}^{D} + (2-\epsilon) B_{22}^{D}) \right. \\
&+& \left. (Q_D \rightarrow -Q_U, D \rightarrow U, \tilde{D}
    \rightarrow \tilde{U}, \theta_D \rightarrow \theta_U)
    \right\}
~~~~~~~~~~~~~~~~(3.2.2)
\end{eqnarray*}
where $\epsilon=4-d$ and d is the space-time dimension.
In the two equations above, we denote
$\{B_{i}^{q},B_{ij}^{q}\}=\{B_{i},B_{ij}\}(p, m_{q}, m_{q}),~
(q=u,d,c,s,t,b)$. $\theta_q$'s as the mixing angles of squark sectors,
respectively. Then the renormalization conditions yield the corresponding
counterterms of the wave functions and the electric charge as:
$$
\delta Z_{AA}=-\frac{\partial \Sigma_{T}^{AA}(p^2)}{\partial p^2}|_{p^2=0},
~~~~~~~~~~~~~~(3.2.3)
$$
$$
\delta Z_{ZA}=2 \frac{\Sigma_{T}^{AZ}(0)}{m_{Z}^2}=0,
~~~~~~~~~~~~~~~~~~~~(3.2.4)
$$
$$
\delta Z_{e}=\frac{1}{2}\frac{\partial \Sigma_{T}^{AA}(p^2)}{\partial p^2}|
         _{p^2=0}-\frac{s_W}{c_W} \frac{\Sigma_{T}^{AZ}(0)}{m_{Z}^2}=
         -\frac{1}{2}\delta Z_{AA}.
~~~~~~~~(3.2.5)
$$
\par
The chargino wave function corrections $\delta Z_{\tilde{\chi}^{+}_{i}
\tilde{\chi}^{+}_{i}}$'s are determined
in terms of the one-particle irreducible two-point function $i\Gamma_{ii}(p^2)$
for charginos in the DR mass basis. It should be written as\cite{a12}:
$$
\begin{array} {lll}
\Gamma_{ii}(p^2) &=&
         (\rlap/p-m_{t}) +  \left [ \rlap/p P_{L}
    \Sigma^{L}_{ii}(p^2)
   + \rlap/p P_{R} \Sigma^{R}_{ii}(p^2)
   + P_{L} \Sigma^{S,L}_{ii}(p^2)
   + P_{R} \Sigma^{S,R}_{ii}(p^2) \right].
\end{array}
\eqno{(3.2.6)}
$$
\par
  With the Feynman rules of the interactions of quark-squark-chargino
in Eqs.(2.18.1$\sim$4), Eqs.(2.19.1$\sim$4) and Eqs.(2.20.1$\sim$4),
the corresponding unrenormalized chargino self-energies
including CP violation phases read(see Fig.1(g.1))
\begin{eqnarray*}
\Sigma^{S,L}_{\tilde{\chi}^{+}_i\tilde{\chi}^{+}_i}(p^2) &=&
   \frac{- N_c} {16 \pi^2} \sum_{(U,D)=(u,d)}^{(c,s),(t,b)} \sum_{k=1,2} \left\{
  m_D V_{D\tilde{U}_{k}\tilde{\chi}^{+}_{i}}^{(1)}
  V_{D\tilde{U}_{k}\tilde{\chi}^{+}_{i}}^{(2)\ast}
  B_{0}[p,m_D,m_{\tilde{U}_{k}}]  \right. \\
&-&\left.  m_U V_{U\tilde{D}_{k}\tilde{\chi}^{+}_{i}}^{(1)}
  V_{U\tilde{D}_{k}\tilde{\chi}^{+}_{i}}^{(2)\ast}
  B_{0}[p,m_U,m_{\tilde{D}_{k}}]  \right\},
~~~~~~~~~~~~~~~~(3.2.7)
\end{eqnarray*}
\begin{eqnarray*}
\Sigma^{S,R}_{\tilde{\chi}^{+}_i\tilde{\chi}^{+}_i}(p^2) &=&
   \frac{-N_c}{16 \pi^2} \sum_{(U,D)=(u,d)}^{(c,s),(t,b)} \sum_{k=1,2} \left\{
  m_D V_{D\tilde{U}_{k}\tilde{\chi}^{+}_{i}}^{(2)}
  V_{D\tilde{U}_{k}\tilde{\chi}^{+}_{i}}^{(1)\ast}
  B_{0}[p,m_D,m_{\tilde{U}_{k}}] \right. \\
&-&\left. m_U V_{U\tilde{D}_{k}\tilde{\chi}^{+}_{i}}^{(2)}
  V_{U\tilde{D}_{k}\tilde{\chi}^{+}_{i}}^{(1)\ast}
  B_{0}[p,m_U,m_{\tilde{D}_{k}}] \right\},
~~~~~~~~~~~~~~~(3.2.8)
\end{eqnarray*}
$$
\Sigma^{L}_{\tilde{\chi}^{+}_i\tilde{\chi}^{+}_i}(p^2) =
    \frac{N_c}{16 \pi^2} \sum_{(U,D)=(u,d)}^{(c,s),(t,b)} \sum_{k=1,2} \left\{
  V_{D\tilde{U}_{k}\tilde{\chi}^{+}_{i}}^{(1)}
  V_{D\tilde{U}_{k}\tilde{\chi}^{+}_{i}}^{(1)\ast}
  B_{1}[p,m_D,m_{\tilde{U}_{k}}] -
  V_{U\tilde{D}_{k}\tilde{\chi}^{+}_{i}}^{(1)}
  V_{U\tilde{D}_{k}\tilde{\chi}^{+}_{i}}^{(1)\ast}
  B_{1}[p,m_U,m_{\tilde{D}_{k}}] \right\},
\eqno{(3.2.9)}
$$
$$
\Sigma^{R}_{\tilde{\chi}^{+}_i\tilde{\chi}^{+}_i}(p^2) =
     \frac{N_c}{16 \pi^2} \sum_{(U,D)=(u,d)}^{(c,s),(t,b)} \sum_{k=1,2} \left\{
  V_{D\tilde{U}_{k}\tilde{\chi}^{+}_{i}}^{(2)}
  V_{D\tilde{U}_{k}\tilde{\chi}^{+}_{i}}^{(2)\ast}
  B_{1}[p,m_D,m_{\tilde{U}_{k}}] -
  V_{U\tilde{D}_{k}\tilde{\chi}^{+}_{i}}^{(2)}
  V_{U\tilde{D}_{k}\tilde{\chi}^{+}_{i}}^{(2)\ast}
  B_{1}[p,m_U,m_{\tilde{D}_{k}}] \right\},
\eqno{(3.2.10)}
$$
\par
Imposing the on-shell renormalization conditions given in Ref.\cite{se}
\cite{a12}, one can obtain the renormalization constants for the
renormalized $\tilde{\chi}_{1}^{+}$ self-energies as\cite{Han1}:
$$
\delta \Sigma_{\tilde{\chi}^{+}_1\tilde{\chi}^{+}_1}(p^2) = C_{L} \rlap/p P_{L} +
   C_{R} \rlap/p P_{R} - C^{-}_{S} P_{L} - C^{+}_{S} P_{R},
\eqno{(3.2.11)}
$$
From Eq.$(3.2.1)~ \sim ~(3.2.5)$ we find that the self-energies of
$\gamma\gamma$ and $\gamma Z^0$ have no contribution to the relevant
counterterms of the $\gamma\tilde{\chi}^{+}_1 \tilde{\chi}^{+}_1$ vertex.
The renormalization constant for the $\Gamma^{\mu}_{\gamma\tilde{\chi}^{+}_1
\tilde{\chi}^{+}_1}$ vertex is written as:
$$
 \delta \Gamma^{\mu}_{\gamma\tilde{\chi}^{+}_1\tilde{\chi}^{+}_1} =
   -i e \gamma^{\mu}[ C^{L} P_{L} + C^{R} P_{R} ].
\eqno{(3.2.12)}
$$
where
$$
\begin{array} {lll}
C_L &=& \frac{1}{2} (\delta Z^L_{\tilde{\chi}^{+}_i\tilde{\chi}^{+}_i} +
   \delta Z^{L\dag}_{\tilde{\chi}^{+}_i\tilde{\chi}^{+}_i}), \\
C_R &=& \frac{1}{2} (\delta Z^R_{\tilde{\chi}^{+}_i\tilde{\chi}^{+}_i} +
   \delta Z^{R\dag}_{\tilde{\chi}^{+}_i\tilde{\chi}^{+}_i}), \\
C^{-}_{S} &=& \frac{m_{\tilde{\chi}^{+}_i}}{2}
(\delta Z^L_{\tilde{\chi}^{+}_i\tilde{\chi}^{+}_i} +
   \delta Z^{R\dag}_{\tilde{\chi}^{+}_i\tilde{\chi}^{+}_i}) +
   \delta m_{\tilde{\chi}^{+}_i}, \\
C^{+}_{S} &=& \frac{m_{\tilde{\chi}^{+}_i}}{2}
   (\delta Z^R_{\tilde{\chi}^{+}_i\tilde{\chi}^{+}_i} +
   \delta Z^{L\dag}_{\tilde{\chi}^{+}_i\tilde{\chi}^{+}_i}) +
   \delta m_{\tilde{\chi}^{+}_i}.
\end{array}
\eqno{(3.2.13)}
$$
\begin{eqnarray*}
\delta m_{\tilde{\chi}^{+}_i} &=& \frac{1}{2} \tilde{Re}
  \left [ m_{\tilde{\chi}^{+}_{i}} \Sigma^{L}_{\tilde{\chi}^{+}_i\tilde{\chi}^{+}_i}
  (m_{\tilde{\chi}^{+}_{i}}^{2}) + m_{\tilde{\chi}^{+}_{i}}
  \Sigma^{R}_{\tilde{\chi}^{+}_i\tilde{\chi}^{+}_i}(m_{\tilde{\chi}^{+}_{i}}^{2}) +
  \Sigma^{S,L}_{\tilde{\chi}^{+}_i\tilde{\chi}^{+}_i}(m_{\tilde{\chi}^{+}_{i}}^{2}) +
  \Sigma^{S,R}_{\tilde{\chi}^{+}_i\tilde{\chi}^{+}_i}(m_{\tilde{\chi}^{+}_{i}}^{2})
  \right ],~~~~~(3.2.14)
\end{eqnarray*}
\begin{eqnarray*}
\delta Z^{L}_{\tilde{\chi}^{+}_i\tilde{\chi}^{+}_i} &=&
- \tilde{Re}\Sigma^{L}_{\tilde{\chi}^{+}_i\tilde{\chi}^{+}_i}
  (m_{\tilde{\chi}^{+}_{i}}^{2}) - \frac{1}{m_{\tilde{\chi}^{+}_{i}}} \tilde{Re}
  \left [ \Sigma^{S,R}_{\tilde{\chi}^{+}_i\tilde{\chi}^{+}_i} (m_{\tilde{\chi}^{+}_{i}}^2)
  - \Sigma^{S,L}_{\tilde{\chi}^{+}_i\tilde{\chi}^{+}_i} (m_{\tilde{\chi}^{+}_{i}}^2)
  \right ] \\
&-& m_{\tilde{\chi}^{+}_{i}} \frac{\partial}{\partial p^2} \tilde{Re}
  \left \{ m_{\tilde{\chi}^{+}_{i}}\Sigma^{L}_{\tilde{\chi}^{+}_i\tilde{\chi}^{+}_i}(p^2)
  + m_{\tilde{\chi}^{+}_{i}} \Sigma^{R}_{\tilde{\chi}^{+}_i\tilde{\chi}^{+}_i}(p^2)
  \right.\\
&+& \left.  \Sigma^{S,L}_{\tilde{\chi}^{+}_i\tilde{\chi}^{+}_i}(p^2) +
     \Sigma^{S,R}_{\tilde{\chi}^{+}_i\tilde{\chi}^{+}_i}(p^2) \right\} |_
     {p^2=m_{\tilde{\chi}^{+}_{i}}^2},
~~~~~~(3.2.15)
\end{eqnarray*}
\begin{eqnarray*}
\delta Z^{R}_{\tilde{\chi}^{+}_i\tilde{\chi}^{+}_i} &=& -\tilde{Re}\Sigma^{R}_
{\tilde{\chi}_i\tilde{\chi}^{+}_i} (m_{\tilde{\chi}^{+}_{i}}^{2}) -
  m_{\tilde{\chi}^{+}_{i}} \frac{\partial} {\partial p^2} \tilde{Re}
  \left\{ m_{\tilde{\chi}^{+}_{i}} \Sigma^{L}_{\tilde{\chi}^{+}_i\tilde{\chi}^{+}_i}(p^2) +
  m_{\tilde{\chi}^{+}_{i}} \Sigma^{R}_{\tilde{\chi}^{+}_i\tilde{\chi}^{+}_i}(p^2) \right. \\
&+& \left. \Sigma^{S,L}_{\tilde{\chi}^{+}_i\tilde{\chi}^{+}_i}(p^2) +
    \Sigma^{S,R}_{\tilde{\chi}^{+}_i\tilde{\chi}^{+}_i}(p^2) \right \} |_
    {p^2=m_{\tilde{\chi}^{+}_{i}}^2}, ~~~~~~~(3.2.16)
\end{eqnarray*}
where $\tilde{Re}$ takes the real part of the loop integrals. It ensures
the reality of the renormalized Lagrangian.

\par
\begin{center} {\bf 3.3 Renormalized one-loop corrections.} \end{center}
\par
    The renormalized one-loop matrix element involves the contributions from
all the one-loop vertex, box, triangle, quartic and self-energy interaction
diagrams(shown in Fig.1) and their relevant counterterms. We can neglect
some of the s-channel Feynman diagrams shown in Fig.1(e.2), in which each one
involves a quark loop with the exchanging of $\gamma$ or $Z^0$ boson.
This is the consequence of Furry theorem, since the Furry theorem forbids the
production of the spin-one components of the $Z^{0}$ and $\gamma$, and the
contribution from the spin-zero component of the $Z^{0}$ vector boson coupling
with a pair of chargino is very small and can be neglected. The calculation
also shows the $\gamma$ and $Z^0$ exchanging s-channel diagrams in Fig.1(d.2)
and Fig.1(e.1) with a squark loop have no contribution to cross section.
The contribution from each of the $\gamma$ and $Z^0$ exchanging s-channel
diagrams in Fig.1(e.1) is canceled out by the corresponding one with
exchanging incoming photons. Considering only the form factors in matrix
element which contribute to the total cross section, the renormalized
amplitude $\delta{\cal M}_{1-loop}$ can be written
in the following form according to their Lorentz invariant structure:
\begin{eqnarray*}
\delta {\cal M}_{1-loop} & = &
    {\cal M}^{v}+{\cal M}^{b}+{\cal M}^{tr}+{\cal M}^{q}+{\cal M}^{s} \\
&=& {\cal M}^{v,\hat{t}}+{\cal M}^{v,\hat{u}}+{\cal M}^{b,\hat{t}}+
    {\cal M}^{b,\hat{u}}+{\cal M}^{tr,\hat{t}}+{\cal M}^{tr,\hat{u}}+
    {\cal M}^{q}+{\cal M}^{s,\hat{t}}+{\cal M}^{s,\hat{u}} \\
&=& N_{c} \epsilon^{\mu}(p_3)\epsilon^{\nu}(p_4) \bar{u}(p_1) \left\{
   f_{1} \gamma_{\mu}\gamma_{\nu} + f_{2} \gamma_{\nu}\gamma_{\mu} +
   f_{3} \gamma_{\mu}p_{1\nu} + f_{4} \gamma_{\mu}p_{2\nu} \right. \\
&+& \left. f_{5} \gamma_{\nu}p_{1\mu} + f_{6} \gamma_{\nu}p_{2\mu} +
   f_{7}  p_{1\mu}p_{1\nu} + f_{8}  p_{1\mu}p_{2\nu} +
   f_{9}  p_{1\nu}p_{2\mu} \right. \\
&+& \left. f_{10} p_{2\mu}p_{2\nu} +
   f_{11} \rlap/{p}_{3} \gamma_{\mu} \gamma_{\nu} +
   f_{12} \rlap/{p}_{3} \gamma_{\nu} \gamma_{\mu} +
   f_{13} \rlap/{p}_{3} \gamma_{\mu} p_{1\nu} +
   f_{14} \rlap/{p}_{3} \gamma_{\mu} p_{2\nu} \right.  \\
&+& \left. f_{15} \rlap/{p}_{3} \gamma_{\nu} p_{1\mu} +
   f_{16} \rlap/{p}_{3} \gamma_{\nu} p_{2\mu} +
   f_{17} \rlap/{p}_{3} p_{1\mu} p_{1\nu} +
   f_{18} \rlap/{p}_{3} p_{1\mu} p_{2\nu} \right.  \\
&+& \left. f_{19} \rlap/{p}_{3} p_{1\nu} p_{2\mu} +
   f_{20} \rlap/{p}_{3} p_{2\mu} p_{2\nu} +
   f_{21} \gamma_5 \epsilon_{\mu\nu\alpha\beta} p_{1}^{\alpha} p_{3}^{\beta}
   \right.  \\
&+& \left. f_{22} \gamma_5 \epsilon_{\mu\nu\alpha\beta} p_{2}^{\alpha}
   p_{3}^{\beta} \right\} v(p_2),
~~~~~~~(3.3.1)
\end{eqnarray*}
with
$$
f_i = f_{i}^{v}+f_{i}^{b}+f_{i}^{tr}+f_{i}^{q}+f_{i}^{s}~~~
(i=1 \sim 22),
\eqno{(3.3.2)}
$$
where ${\cal M}^{v}$, ${\cal M}^{b}$, ${\cal M}^{tr}$, ${\cal M}^{q}$ and
${\cal M}^{s}$ are the renormalized matrix elements contributed by vertex,
box, triangle, quartic interaction and self-energy corrections, respectively.
$f_{i}^{v}$, $f_{i}^{b}$, $f_{i}^{tr}$, $f_{i}^{q}$ and $f_{i}^{s}$ are the
corresponding form factors. We divide the matrix elements ${\cal M}^{v}$,
${\cal M}^{b}$, ${\cal M}^{tr}$ and ${\cal M}^{s}$ into t- and u-channel
parts, respectively. For each of the corresponding form factor we have
$$
f_{i}^{v}=f_{i}^{v,\hat{t}}+f_{i}^{v,\hat{u}},~
f_{i}^{b}=f_{i}^{b,\hat{t}}+f_{i}^{b,\hat{u}},~
f_{i}^{tr}=f_{i}^{tr,\hat{t}}+f_{i}^{tr,\hat{u}},~
f_{i}^{s}=f_{i}^{s,\hat{t}}+f_{i}^{s,\hat{u}}.
(i=1 \sim 22),
$$
Since the amplitude parts from the u-channel vertex, box and quartic
interaction corrections can be obtained from the t-channel's by doing
exchange as below:
\begin{eqnarray*}
{\cal M}^{j,\hat{u}}={\cal M}^{j,\hat{t}}(t \rightarrow u,
   p_3 \leftrightarrow p_4, \mu \leftrightarrow \nu),~~(j=v,b,tr,s)
\end{eqnarray*}
we list only the explicit t-channel form factors $f_i^{v,\hat{t}}$,
$f_i^{b,\hat{t}}$ and $f_i^{tr,\hat{t}}~(i=1\sim 22)$ in Appendix.
Then we can arrive at the (s)quarks one-loop corrections of the
cross section for this subprocess in unpolarized photon collisions.
\begin{eqnarray*}
 \delta \hat{\sigma}^{1-loop}(\hat{s}) &=& \frac{1}{16 \pi \hat{s}^2}
             \int_{\hat{t}^{-}}^{\hat{t}^{+}} d\hat{t}~
        2 Re {\bar{\sum\limits_{spins}^{}}} \left( {\cal M}_{0}^{\dag} \cdot
        \delta {\cal M}_{1-loop} \right),~~~~~~~(3.3.3)
\end{eqnarray*}
where $\hat{t}^\pm=(m_{\tilde{\chi}_1}^2-\frac{1}{2}\hat{s})\pm\frac{1}{2}\hat{s}
\beta$. The bar over the sum means that we are taking average over initial
spins.
\par
In our calculation, some parts of self-energies and vertex corrections are
similar with those in \cite{Diaz}\cite{singo}. The comparison with their
expressions have been performed. We find that our self-energy of chargino
agrees with the expressions in References \cite{Diaz} and \cite{Singo}
when there is no CP-violation phases, but the self-energies of $\gamma$
and $\gamma-Z^0$ mixing from quark loops presented above have discrepancies
with equations (C.11) and (C.13) in Ref.\cite{Diaz}. The self-energy
parts from quark loop in Eqs.(3.2.1) and (3.2.2) in our paper are coincident
with the expressions for the SM\cite{Denner}.

\par
\begin{flushleft} {\bf 4. Numerical results and discussions} \end{flushleft}

\par
In the following numerical evaluation, we present some results of the
one-loop radiative corrections of virtual (s)quarks to the cross sections
of the lightest chargino pair production in the processes of
$\gamma \gamma \rightarrow \tilde{\chi}^{+}_{1} \tilde{\chi}^{-}_{1}$ and
$e^+ e^- \rightarrow \gamma \gamma \rightarrow \tilde{\chi}^{+}_{1}
\tilde{\chi}^{-}_{1}$, respectively. The input parameters on the MSSM can be
divided into two parts. One is for the SM parameters and the other is
for SUSY parameters. The SM parameters are chosen as:
$m_t=175~GeV$, $m_{Z}=91.187~GeV$, $m_b=4.5~GeV$,
$\sin^2{\theta_{W}}=0.2315$, and $\alpha = 1/137.036$.
The SUSY parameters are taken as follows by default unless otherwise stated.
\par
(1) The masses of squark mass eigenstates $\tilde{U}_{1,2}~
(\tilde{U}=\tilde{u},\tilde{c},\tilde{t})$ and $\tilde{D}_{1,2}~(\tilde{D}=
\tilde{d},\tilde{s},\tilde{b})$ are determined by Eqs.$(2.2~\sim~2.10)$.
From renormalization group equations \cite{Drees2} one expects that the
soft SUSY breaking masses $m_{\tilde{q}_{L}}$ and $m_{\tilde{q}_{R}}$ of
the third generation squarks are smaller than those of the first and
second generations due to the Yukawa interactions. The mixing between
the left- and right-handed stop quarks $\tilde{t}_{L}$ and $\tilde{t}_{R}$
can be very large due to the large mass of the top quark, and the lightest
scalar top quark mass eigenstate $\tilde{t}_1$ can be much lighter than the
top mass and all the scalar partners of the light quarks. Here the left-right
mixing of the top squark plays an important role.
We assume $m_{\tilde{U}_1} < m_{\tilde{U}_2}$, $m_{\tilde{D}_1} <
m_{\tilde{D}_2}$ and for simplicity we take the stop and sbottom
mixing angles being the values of $\theta_{t}= \frac{\pi}{4}$ and
$\theta_{b}=0$, respectively, but the mixing angles of the first and second
generation squarks are set to be $\tan{\theta_{u,d,c,s}}=0.2$. In numerical
calculation, we take $\tilde{M}_Q = \tilde{M}_U = \tilde{M}_D = \tilde{M} =
200~GeV$ (for the third generation) and $600~GeV$ (for the first and second
generations). The ratio of the vacuum expectation values $\tan{\beta}$ is
set to be 4 or 40 in order to make comparison.
\par
(2) The neutral Higgs boson masses $m_{h^0}$, $m_{H^0}$ and $m_{A^0}$ are
given by\cite{Esp}
\begin{eqnarray*}
m_{h^0,H^0}^2 &=& \frac{1}{2}\left[ m_{A^0}^2+m_{Z}^2+\epsilon \mp  \right. \\
  && \left. \sqrt{(m_{A^0}^2+m_{Z}^2+\epsilon)^2-4m_{A^0}^2m_{Z}^2
  cos^2{2\beta}- 4\epsilon(m_{A^0}^2sin^2{\beta}+m_{Z}^2cos^2{\beta}) } \right],
~~~~~~~~~~(4.1)
\end{eqnarray*}
$$
m_{H^{\pm}}^2= m_{A^0}^2+m_{W}^2
~~~~~~~~~~~(4.2)
$$
with the leading corrections being characterized by the radiative parameter
$\epsilon$
$$
\epsilon = \frac{3 G_{F}}{\sqrt{2}\pi^2} \frac{m_t^4}{sin^2{\beta}} log \left[
         \frac{m_{\tilde{t}}^2}{m_{t}^2} \right] .
~~~~~~~~~~~(4.3)
$$
The parameter $m_{\tilde{t}}^2=m_{\tilde{t}_1}m_{\tilde{t}_2}$ denotes
the average squared mass of the stop quarks. The mixing angle $\alpha$
is fixed by $\tan{\beta}$ and the Higgs boson mass $m_{A^0}$,
\begin{eqnarray*}
\tan{2 \alpha}=\tan{2 \beta} \frac{m_{A^0}^2+m_{Z}^2} {m_{A^0}^2-m_{Z}^2
               + \frac{\epsilon}{\cos{2 \beta}} }~~(-\frac{\pi}{2}<\alpha<0).
~~~(4.4)
\end{eqnarray*}
In this paper, we take $m_{A^0}=150~GeV$.
\par
(3) The physical chargino masses $m_{\tilde{\chi}^{+}_{1}}$ and
$m_{\tilde{\chi}^{+}_{2}}$ are set to be
165 GeV and 750 GeV, respectively. Assuming that $\mu$ has positive sign,
the fundamental SUSY parameters $M_{SU(2)}$ and $|\mu|$ can be extracted at
the tree level from these input chargino masses, $\tan{\beta}$ and the
complex phase angle of $\mu$ by using Eqs.(2.15). When the lightest chargino
is dominantly gaugino (gaugino-like or wino-like), we should have
$M_{SU(2)} << |\mu|$ from Eq.(2.15), and when the chargino is dominantly
Higgsino (Higgsino-like), $M_{SU(2)}$ should be much larger than $|\mu|$.
Here we define the relative corrections for subprocess
$\gamma\gamma \rightarrow \tilde{\chi}_1^{+} \tilde{\chi}_1^{-}$ and
process $e^+ e^- \rightarrow \gamma\gamma \rightarrow \tilde{\chi}_1^{+}
\tilde{\chi}_1^{-}$ as
$$
\hat{\delta}=\frac{\delta\hat{\sigma}^{1-loop}}{\hat{\sigma}_{0}},~~
\delta=\frac{\delta\sigma^{1-loop}}{\sigma_{0}},~~~~(4.5)
$$
respectively. The relative corrections of all the quarks and their SUSY
partners as the functions of the c.m.s. energy of photons $\sqrt{\hat{s}}$
with $m_{\tilde{\chi}^{+}_{1}}=165~GeV$, $m_{\tilde{\chi}^{+}_{2}}=750~GeV$ and
all CP phases being zero, are shown in Fig.3(a) and (b). In figure 3(a) the two
curves correspond to the Higgsino-like and gaugino-like chargino pair
productions with $\tan{\beta}=4$, respectively, while the plot in Figure 3(b)
is for both Higgsino-like and
gaugino-like chargino cases with $\tan{\beta}=40$. In general, the corrections
with $\tan{\beta}=40$ are approximately twice as large as those with
$\tan{\beta}=4$. Because of the resonance effects, all the four curves in
Fig.3(a) $\sim$ (b) have peaks or spikes at the energy positions where
the resonance conditions are satisfied. The large enhancement peaks on all
four curves are located in the vicinity of $\sqrt{\hat{s}}=2 m_{t}=350~GeV$.
On the two curves of Fig.3(a) with $\tan{\beta}=4$, there are two small spikes
stemming from resonance effects in the vicinities of $\sqrt{\hat{s}} \sim
2 m_{\tilde{b}_{1,2}} \sim 403~GeV$ and $\sqrt{\hat{s}}=
2 m_{\tilde{t}_2} \sim 674~GeV$, whereas for the two curves in Fig.3(b)
with $\tan{\beta}=40$, the small spikes due to resonance effect are
located at the positions of $\sqrt{\hat{s}} \sim 2 m_{\tilde{b}_{1,2}} \sim
400~GeV$ and $\sqrt{\hat{s}}=2 m_{\tilde{t}_2}\sim 690~GeV$, respectively.
Figure 4 gives the relative corrections of the subprocess as a function of
self-supersymmetry-breaking mass parameter of the third generation scalar
quarks $\tilde{M}$ (We set the $\tilde{M}$ for the first and second generations
being 600 GeV.) with $\sqrt{\hat{s}}=400~GeV$ and all three CP-odd phases
being zero. In this figure we can see considerable enhancement around the
points of $\tilde{M}=205~GeV$ for $\tan{\beta}=4$ and $\tilde{M}=219~GeV$
for $\tan{\beta}=40$, due to the influence of the singularity in the lightest
chargino wave function renormalization. This singularity originates from
the renormalization constants $\delta Z^{L}_{\tilde{\chi}^{+}_{1}
\tilde{\chi}^{+}_{1}}$ and $\delta Z^{R}_{\tilde{\chi}^{+}_{1}
\tilde{\chi}^{+}_{1}}$ of the lightest chargino wave function (see Eq.(3.2.9)
and Eq.(3.2.10)) at the point in parameter space where
$m_{\tilde{\chi}^{+}_{1}} = m_{\tilde{t}_1}+ m_{b}$. And in the
vicinities of $\tilde{M} \sim 235~GeV$ (for $\tan{\beta}=4$ curves) and
$\tilde{M} \sim 250~GeV$ (for $\tan{\beta}=40$ curves),
there is a small suppression spike which shows the resonance effect on each
curve, where $\sqrt{\hat{s}}=400~GeV \approx 2 m_{\tilde{t}_1}$.
We can also see that in the Higgsino-like case the relative corrections for
$\tan{\beta}=40$ are generally larger than those for $\tan{\beta}=4$,
but in gaugino-like case the dependence of the correction on the $\tan{\beta}$
is not so clear except in the vicinity of the singularity region. All
the four curves show that the relative corrections have weak dependence on
$\tilde{M}$ except in some special regions.

The relative correction of the subprocess cross section versus the
lightest chargino mass $m_{\tilde{\chi}^{+}_{1}}$ with $\sqrt{\hat{s}}=400~GeV$
and $\phi_{\mu}=\phi_{t}=\phi_{b}=0$, are depicted in figure 5. The four
curves show obvious correction enhancement at the positions of
$m_{\tilde{\chi}^{+}_{1}}=160~GeV$ for $\tan{\beta}=4$ and
$m_{\tilde{\chi}^{+}_{1}}=139~GeV$ for $\tan{\beta}=40$, respectively.
This is once again the effect of the singularity due to the renormalization
constants in the lightest chargino wave function when
$m_{\tilde{\chi}^{+}_{1}} = m_{\tilde{t}_1}+ m_{b}$.
\par
The relative corrections in the subprocess of Higgsino-like chargino pair
production versus the CP phases angles $\phi_{CP}(\phi_{\mu},\phi_{t,b})$
with $\sqrt{\hat{s}}=400~GeV$, $m_{\tilde{\chi}^{+}_1}=165~GeV$ and
$m_{\tilde{\chi}^{+}_2}=750~GeV$, are depicted in figure 6. The full-line
and dotted-line correspond to $\phi_{CP}=\phi_t=\phi_b$ with $\tan{\beta}=4$
and $\tan{\beta}=40$, respectively. The dashed-line and dash-and-dotted-line
correspond to $\phi_{CP}=\phi_{\mu}$ with $\tan{\beta}=4$ and $\tan{\beta}=40$,
respectively. Fig.6 shows the periodical features of
$\hat{\delta}(\phi_t)=\hat{\delta}(\pi+\phi_t)$ for the curves of
$\hat{\delta}$ versus $\phi_t$ and
$\hat{\delta}(\phi_{\mu})=\hat{\delta}(2 \pi+\phi_{\mu})$ for the curves of
$\hat{\delta}$ versus $\phi_{\mu}$, respectively. All the three CP phase
angles affect the corrections obviously.
\par
From our numerical calculation with the parameters chosen above, we find in
the Higgsino-like chargino pair production the
radiative corrections from the first and second generation quarks and squarks
are only few millesimal of the total corrections for our choice of parameters.
The corrections are mainly from the loop diagrams of
top, bottom quarks and their supersymmetric partners.
But in the gaugino-like case, the radiative corrections from the first and
second generation quarks and squarks can reach 3 percent of the total
contribution.
\par
The chargino pair production via photon-photon fusion
is only a subprocess of the parent $e^{+}e^{-}$ linear collider. It is easy
to obtain the total cross section of the lightest chargino pair production via
photon fusion in $e^{+}e^{-}$ collider, by folding the cross
section of the subprocess $\hat{\sigma} (\gamma\gamma \rightarrow
\tilde{\chi}^{+}_1 \bar{\tilde{\chi}}^{-}_1)$ with the photon luminosity.
$$
\sigma(s)=\int_{2 m_{\tilde{t}_1}/\sqrt{s} }^{x_{max}}
dz \frac{dL_{\gamma\gamma}}{dz} \hat{\sigma}(\gamma\gamma\rightarrow
 \tilde{\chi}^{+}_1 \bar{\tilde{\chi}}^{-}_1\hskip 3mm at \hskip 3mm \hat{s}=z^2 s),
\eqno{(4.6)}
$$
where $\sqrt{s}$ and $\sqrt{\hat{s}}$ are the $e^{+}e^{-}$ and $\gamma \gamma$
c.m.s. energies respectively and $dL_{\gamma\gamma}/dz$
is the distribution function of photon luminosity, which is
$$
\frac{dL_{\gamma\gamma}}{dz}=2z\int_{z^2/x_{max}}^{x_{max}}
 \frac{dx}{x} f_{\gamma/e}(x)f_{\gamma/e}(z^2/x),
\eqno{(4.7)}
$$
where $f_{\gamma/e}$ is the photon structure function of the electron beam
\cite{sd,sh}. We take the structure function of the photon produced by
Compton backscattering as \cite{sd,si}
\[
f^{Comp}_{\gamma/e} =\left\{
\begin{array}{cl}
\frac{1}{1.8397} \left(1-x+\frac{1}{1-x}-\frac{4x}{x_{i}(1-x)}+\frac{4 x^2}{x_{i}^2 (1-x)^2} \right),
& {\rm for}~x~<~0.83, x_{i}=2(1+\sqrt{2}) \\
0, & {\rm for}~x~>~0.83.
\end{array}
\right.
\]
$$
\eqno{(4.8)}
$$
The cross section including all (s)quark loop corrections for
the process of the lightest Higgsino-like chargino pair production
$e^{+}e^{-} \rightarrow \gamma \gamma \rightarrow \tilde{\chi}^{+}_1
\bar{\tilde{\chi}^{-}_1}$ versus $\sqrt{s}$, with $m_{\tilde{\chi}^{+}_{1}}=
165~GeV$, $m_{\tilde{\chi}^{+}_{2}}=750~GeV$ and all of CP phases being zero,
are depicted in Fig.7(a). Their relative corrections $\delta$ versus $\sqrt{s}$
are presented in Fig.7(b). By analysing our numerical data, we can see the
value of the relative correction $\delta$ can approach $3.2\%$ in
Higgsino-like chargino pair production, when $\sqrt{s}$ is
$0.5~TeV$, $\tan{\beta}=40$ and all CP phase angles are set to zero.
Fig.7(a) shows that the cross section of the Higgsino-like chargino pair
production via photon-photon collision in the NLC can be over one pico-bar.
In Fig.7(b) we can see that the relative corrections have very
weak dependence on the c.m.s energy $\sqrt{s}$ both for the Higgsino-like
and gaugino-like chargino cases, when the c.m.s energy
$\sqrt{s}$ is beyond 900 GeV.
\par
After comparing our numerical results with those in References \cite{Singo}
and \cite{Diaz} which deal with the chargino pair production via $e^+ e^-$
collision, we find the numerical order of the corrections in this paper
approach the former \cite{Singo} instead of the latter \cite{Diaz}, though
the processes investigated are different. Moreover, Our calculation shows
that the contributions from (s)quark loops from all the three generations
should be considered, though generally the contribution from the third
generation dominates, but in gaugino-like chargino case the contributions
of the first two generations cannot be neglected in precise calculation.

\par
\begin{flushleft} {\bf 5. Summary} \end{flushleft}
\par
In this paper, we have analysed the virtual one-loop corrections of
all the (s)quarks within the MSSM to the lightest chargino pair
production at the future NLC operating in photon-photon collision mode.
From the results of numerical evaluation for several typical parameter
sets, it shows that the radiative corrections arising from all virtual
quarks and their supersymmetric partners enhance the cross sections
significantly, especially in the Higgsino-like chargino pair production
process. The corrections parts contributed from the first and second
generation (s)quarks are few millesimal of the total for Higgsino-like case
and less than $3\%$ for gaugino-like case, respectively. Typically, the Born
cross sections for subprocess are enhanced by about one or two percent in
Higgsino-like chargino case, but less than $1\%$ in gaugino-like chargino case.
In some exceptional c.m.s energy regions, where the resonance conditions are
satisfied or the singularity point of the wave function renormalization
constants of $\tilde{\chi}^{+}_1$ exists in the parameter space, the relative
corrections may be observably enhanced even by three or four percent. We also
investigate the effects of complex phases $\phi_{t,b}$ in the squark mass
matrices and $\phi_{\mu}$ appearing in chargino mass matrix in Higgsino-like
case. We find the radiative corrections are very sensitive to the CP-odd
complex phases $\phi_{\mu}$ and $\phi_{t,b}$. Since the lightest chargino mass,
its pair production cross section and radiative corrections in both subprocess
and parent process, are all strongly related to the CP-odd complex phase
$\phi_{\mu}$, the experimental determination of the
parameters $\phi_{\mu}$ and $\phi_{t,b}$ is crucial in searching for
SUSY signals.

\vskip 4mm
\noindent{\large\bf Acknowledgement:}
These work was supported in part by the National Natural Science
Foundation of China(project numbers: 19675033, 19875049), the Youth Science
Foundation of the University of Science and Technology of China and
a grant from the State Commission of Science and Technology of China.

\vskip 5mm
\begin{center} {\Large Appendix}\end{center}

In this appendix we list only the form factors for the third generation
quarks and squarks, and in fact we should take the sum of the form factors
of the three generations for the total form factors.  Some of the results
have been cross-checked and they fit well with other literature such as Ref.
\cite{Diaz}.

In the following, we use the notations defined as below.
$$
\bar{B}_{0}^{(1,k)}=B_{0}[-p_1-p_2,m_{\tilde{t}_k},m_{\tilde{t}_k}]-\Delta,~~~
\bar{B}_{0}^{(2,k)}=\bar{B}_{0}^{(1,k)}(m_{\tilde{t}_k} \rightarrow m_{\tilde{b}_k}),
$$
$$
B_{0}^{(3,k)},B_{1}^{(3,k)}=B_{0},B_{1}[p_1-p_3, m_b, m_{\tilde{t}_k}],~~~
B_{0}^{(4,k)},B_{1}^{(4,k)}=B_{0},B_{1}[p_3-p_1, m_t, m_{\tilde{b}_k}],
$$
$$
C_{0}^{(1,k)},C_{ij}^{(1,k)}=
    C_{0},C_{ij}[p_1,-p_1-p_2,m_b,m_{\tilde{t}_k},m_{\tilde{t}_k}],
$$
$$
C_{0}^{(2,k)},C_{ij}^{(2,k)}= C_{0}^{(5,k)},C_{ij}^{(5,k)}
   [m_b,m_{\tilde{t}_k} \rightarrow m_t,m_{\tilde{b}_k}].
$$
$$
C_{0}^{(3,k)},C_{ij}^{(3,k)}=
    C_{0},C_{ij}[p_3,-p_1-p_2,m_{\tilde{t}_k},m_{\tilde{t}_k},m_{\tilde{t}_k}].
$$
$$
C_{0}^{(4,k)},C_{ij}^{(4,k)}= C_{0}^{(3,k)},C_{ij}^{(3,k)}
   (p_1,p_2,p_3,m_{\tilde{t}_k} \rightarrow -p_1,-p_2,-p_3,m_{\tilde{b}_k}).
$$
$$
C_{0}^{(5)},C_{ij}^{(5)}= C_{0},C_{ij}[-p_3,p_1+p_2,m_b,m_b,m_b],~~
C_{0}^{(6)},C_{ij}^{(6)}= C_{0}^{(5)},C_{ij}^{(5)}(m_b \rightarrow m_t).
$$
$$
C_{0}^{(7,k)},C_{ij}^{(7,k)}=
        C_{0},C_{ij}[-k_1,k_1+k_2,m_t,m_{\tilde{b}_k},m_{\tilde{b}_k}],
$$
$$
C_{0}^{(8,k)},C_{ij}^{(8,k)}=
        C_{0},C_{ij}[k_1,-k_1-k_2,m_b,m_{\tilde{t}_k},m_{\tilde{t}_k}],
$$
$$
C_{0}^{(9,k)},C_{ij}^{(9,k)}=
        C_{0},C_{ij}[k_1,-k_1-k_2,m_{\tilde{b}_k}, m_t, m_t],
$$
$$
C_{0}^{(10,k)},C_{ij}^{(10,k)}=
        C_{0},C_{ij}[-k_1,k_1+k_2,m_{\tilde{t}_k},m_b,m_b].
$$

$$
D_{0}^{(1,k)},D_{ij}^{(1,k)},D_{ijl}^{(1,k)}=
    D_{0},D_{ij},D_{ijl}[-p_1,p_3,p_4,m_{\tilde{t}_k},m_b,m_b,m_b]
$$
$$
D_{0}^{(2,k)},D_{ij}^{(2,k)},D_{ijl}^{(2,k)}=
    D_{0},D_{ij},D_{ijl}[p_1,-p_3,-p_4,m_{\tilde{b}_k},m_t,m_t,m_t]
$$
$$
D_{0}^{(3,k)},D_{ij}^{(3,k)},D_{ijl}^{(3,k)}=
    D_{0},D_{ij},D_{ijl}[p_1,-p_3,-p_4,
                m_b,m_{\tilde{t}_k},m_{\tilde{t}_k},m_{\tilde{t}_k}]
$$
$$
D_{0}^{(4,k)},D_{ij}^{(4,k)},D_{ijl}^{(4,k)}=
   D_{0},D_{ij},D_{ijl}[-p_1,p_3,p_4,
                m_t,m_{\tilde{b}_k},m_{\tilde{b}_k},m_{\tilde{b}_k}]
$$
$$
D_{0}^{(5,k)},D_{ij}^{(5,k)},D_{ijl}^{(5,k)}=
    D_{0},D_{ij},D_{ijl}[p_1,-p_3,p_2,m_b,m_{\tilde{t}_k},m_{\tilde{t}_k},m_b]
$$
$$
D_{0}^{(6,k)},D_{ij}^{(6,k)},D_{ijl}^{(6,k)}=
    D_{0},D_{ij},D_{ijl}[-p_1,p_3,-p_2,m_t,m_{\tilde{b}_k},m_{\tilde{b}_k},m_t]
$$
We defined some factors as below:
\begin{eqnarray*}
A_{t}=\frac{i}{\hat{t}-m_{\tilde{\chi}^{+}_1}^2},~~~~
A_{u}=\frac{i}{\hat{u}-m_{\tilde{\chi}^{+}_1}^2},
\end{eqnarray*}
\begin{eqnarray*}
A_{h}=\frac{i}{\hat{s}-m_{h}^2},~~~~~
A_{H}=\frac{i}{\hat{s}-m_{H}^2}.
\end{eqnarray*}
\begin{eqnarray*}
A_{A}=\frac{i}{\hat{s}-m_{A}^2},~~~~~
A_{G}=\frac{i}{\hat{s}-m_{Z}^2}.
\end{eqnarray*}
$$
F_{1}^{q\tilde{q}^{\prime}_{k}\tilde{\chi}^{+}_{1}}=
   |V_{q\tilde{q}^{\prime}_{k}\tilde{\chi}^{+}_{1}}^{(1)}|^2+
   |V_{q\tilde{q}^{\prime}_{k}\tilde{\chi}^{+}_{1}}^{(2)}|^2,~~~
F_{2}^{q\tilde{q}^{\prime}_{k}\tilde{\chi}^{+}_{1}}=
    V_{q\tilde{q}^{\prime}_{k}\tilde{\chi}^{+}_{1}}^{(1)\ast}
    V_{q\tilde{q}^{\prime}_{k}\tilde{\chi}^{+}_{1}}^{(2)}+
    V_{q\tilde{q}^{\prime}_{k}\tilde{\chi}^{+}_{1}}^{(2)\ast}
    V_{q\tilde{q}^{\prime}_{k}\tilde{\chi}^{+}_{1}}^{(1)}.
$$
where $q\tilde{q}^{\prime}_{k}\tilde{\chi}^{+}_{1}=
 t\tilde{b}_{k}\tilde{\chi}^{+}_{1},
~b\tilde{t}_{k}\tilde{\chi}^{+}_{1}$.
In the following, the expression denoted as $(t \rightarrow b)$ means
replacements of $Q_t \rightarrow Q_b$, $m_t \rightarrow m_b$
and $F_{i}^{t\tilde{b}_{k}\tilde{\chi}^{+}_1} \rightarrow
F_{i}^{b\tilde{t}_{k}\tilde{\chi}^{+}_1}$ $(i=1 \sim 2)$.
\par
The one-particle-irreducible(1PI) correction to the vertex of
$\gamma\tilde{\chi}^{+}_1\tilde{\chi}^{+}_1$ from quarks and their
SUSY partners, can be written in terms of form factors.
\begin{eqnarray*}
\Delta\Gamma_{\gamma\tilde{\chi}^{+}_1\tilde{\chi}^{+}_1}^{\mu}(k_1,k_2)&=&
   g_{1}(k_1,k_2) k_{1}^{\mu} \gamma_5 \rlap/{k}_1
   + g_{2}(k_1,k_2) k_{2}^{\mu} \gamma_5 \rlap/{k}_1
   + g_{3}(k_1,k_2) k_{1}^{\mu} \gamma_5 \rlap/{k}_2 \\
&+& g_{4}(k_1,k_2) k_{2}^{\mu} \gamma_5 \rlap/{k}_2
   + g_{5}(k_1,k_2) k_{1}^{\mu} \gamma_5
   + g_{6}(k_1,k_2) k_{2}^{\mu} \gamma_5  \\
&+& g_{7}(k_1,k_2) \gamma_5 \gamma^{\mu} \rlap/{k}_1 \rlap/{k}_2
   + g_{8}(k_1,k_2) \gamma_5 \gamma^{\mu} \rlap/{k}_1
   + g_{9}(k_1,k_2) \gamma_5 \gamma^{\mu} \rlap/{k}_2 \\
&+& g_{10}(k_1,k_2) \gamma_5 \gamma^{\mu}
   + g_{11}(k_1,k_2) k_{1}^{\mu} \rlap/{k}_1
   + g_{12}(k_1,k_2) k_{2}^{\mu} \rlap/{k}_1 \\
&+& g_{13}(k_1,k_2) k_{1}^{\mu} \rlap/{k}_2
   + g_{14}(k_1,k_2) k_{2}^{\mu} \rlap/{k}_2
   + g_{15}(k_1,k_2) k_{1}^{\mu} \\
&+& g_{16}(k_1,k_2) k_{2}^{\mu}
   + g_{17}(k_1,k_2) \gamma^{\mu} \rlap/{k}_1 \rlap/{k}_2
   + g_{18}(k_1,k_2) \gamma^{\mu} \rlap/{k}_1 \\
&+& g_{19}(k_1,k_2) \gamma^{\mu} \rlap/{k}_2
   + g_{20}(k_1,k_2) \gamma^{\mu},
\end{eqnarray*}
where $k_1$ and $k_2$ are the four-momenta of the lightest chargino pair and
in their out going directions, respectively. In the equation above, the form
factors of the Lorentz invariant structures including $\gamma_5$ do not
contribute to the cross sections of our subprocess. Therefore we shall list
only the explicit expressions of the form factors $g_{i}~(i=11 \sim 20)$.
The form factors $g_{i}~(i=11 \sim 20)$ are expressed
explicitly as follows.
\begin{eqnarray*}
g_{11}(k_1, k_2) &=& \frac{-i e}{32 \pi^2} \sum_{k=1,2} \left\{
    F_{1}^{t\tilde{b}_{k}\tilde{\chi}^{+}_{1}}
   \left[ Q_b (C_{11}^{(7,k)}-C_{12}^{(7,k)}+
   2 C_{21}^{(7,k)}+2 C_{22}^{(7,k)}-
   4  C_{23}^{(7,k)}) \right.\right. \\
&+&\left. 2 Q_t (C_{11}^{(9,k)}-C_{12}^{(9,k)}+
   C_{21}^{(9,k)}+C_{22}^{(9,k)}-2 C_{23}^{(9,k)})
   \right] \\
&+& \left. (t,b,C^{(7)}, C^{(9)} \rightarrow b,t,C^{(8)}, C^{(10)} ) \right\},
\end{eqnarray*}
\begin{eqnarray*}
g_{12}(k_1, k_2) &=& \frac{i e}{32 \pi^2} \sum_{k=1,2} \left\{
    F_{1}^{t\tilde{b}_{k}\tilde{\chi}^{+}_{1}}
    \left[ Q_b (C_{11}^{(7,k)}-C_{12}^{(7,k)}
    -2 C_{22}^{(7,k)}+2 C_{23}^{(7,k)} )-
    2 Q_t(C_{22}^{(9,k)}-C_{23}^{(9,k)}) \right.\right]  \\
&+&\left.(t,b,C^{(7)}, C^{(9)} \rightarrow b,t,C^{(8)}, C^{(10)} ) \right\},
\end{eqnarray*}
\begin{eqnarray*}
g_{13}(k_1, k_2) &=& \frac{i e}{32 \pi^2} \sum_{k=1,2}\left\{
    F_{1}^{t\tilde{b}_{k}\tilde{\chi}^{+}_{1}}
   \left[ Q_b (C_{12}^{(7,k)}-2 C_{22}^{(7,k)}+
   2 C_{23}^{(7,k)}) \right.\right] \\
&& \left. +2 Q_t (C_{0}^{(9,k)}+
   C_{11}^{(9,k)}-C_{22}^{(9,k)}+
   C_{23}^{(9,k)} )
  +(t,b,C^{(7)}, C^{(9)} \rightarrow b,t,C^{(8)}, C^{(10)} ) \right\},
\end{eqnarray*}
\begin{eqnarray*}
g_{14}(k_1, k_2) &=& \frac{-i e}{32 \pi^2} \sum_{k=1,2} \left\{
    F_{1}^{t\tilde{b}_{k}\tilde{\chi}^{+}_{1}}
   \left[ Q_b (C_{12}^{(7,k)}+2 C_{22}^{(7,k)})+
   2 Q_t (C_{12}^{(9,k)}+C_{22}^{(9,k)}) \right.\right] \\
&+&\left.(t,b,C^{(7)}, C^{(9)} \rightarrow b,t,C^{(8)}, C^{(10)} ) \right\},
\end{eqnarray*}
\begin{eqnarray*}
g_{15}(k_1, k_2) &=& \frac{i e}{32 \pi^2} \sum_{k=1,2} \left\{ m_t
    F_{2}^{t\tilde{b}_{k}\tilde{\chi}^{+}_{1}}
    \left[ Q_b (C_{0}^{(7,k)}+2 C_{11}^{(7,k)}-
    2 C_{12}^{(7,k)})-2 Q_t (C_{0}^{(9,k)}+
    C_{11}^{(9,k)} \right. \right.\\
&-& \left. \left. C_{12}^{(9,k)}) \right]
    +(t,b,C^{(7)}, C^{(9)} \rightarrow b,t,C^{(8)}, C^{(10)} ) \right\},
\end{eqnarray*}
\begin{eqnarray*}
g_{16}(k_1, k_2) &=& \frac{-i e}{32 \pi^2} \sum_{k=1,2} \left\{ m_t
    F_{2}^{t\tilde{b}_{k}\tilde{\chi}^{+}_{1}}
    \left[ Q_b (C_{0}^{(7,k)}+2 C_{12}^{(7,k)})-
    2 Q_t C_{12}^{(9,k)} \right.\right] \\
&+& \left.(t,b,C^{(7)}, C^{(9)} \rightarrow b,t,C^{(8)}, C^{(10)} ) \right\},
\end{eqnarray*}
\begin{eqnarray*}
g_{17}(k_1, k_2) &=& \frac{-i e}{32 \pi^2} \sum_{k=1,2} \left\{ Q_t
    F_{1}^{t\tilde{b}_{k}\tilde{\chi}^{+}_{1}}
    (C_{0}^{(9,k)}+C_{11}^{(9,k)})
   +(t,b,C^{(9)} \rightarrow b,t,C^{(10)} )   \right\},
\end{eqnarray*}
\begin{eqnarray*}
g_{18}(k_1, k_2) &=& g_{19}(k_1, k_2) =
 \frac{i e}{32 \pi^2} \sum_{k=1,2} \left\{ m_t Q_t
    F_{2}^{t\tilde{b}_{k}\tilde{\chi}^{+}_{1}} C_{0}^{(9,k)}
    +(t,b,C^{(9)} \rightarrow b,t C^{(10)} ) \right\},
\end{eqnarray*}
\begin{eqnarray*}
g_{20}(k_1, k_2) &=& \frac{i e}{32 \pi^2} \sum_{k=1,2} \left\{
    F_{1}^{t\tilde{b}_{k}\tilde{\chi}^{+}_{1}}
   \left[ 2 Q_b C_{24}^{(7,k)}-Q_t ( m_t^2 C_{0}^{(9,k)}-
    k_1^2 C_{11}^{(9,k)}+(k_1^2-k_2^2) C_{12}^{(9,k)}
    \right.\right. \\
&-& \left.\left. k_1^2 C_{21}^{(9,k)}-
    (k_1+k_2)^2 C_{22}^{(9,k)}+
    2 (k_1^2+k_1 \cdot k_2)C_{23}^{(9,k)}+(2-\epsilon) C_{24}^{(9,k)}
     ) \right] \right. \\
&&\left.+(t,b, C^{(7)}, C^{(9)} \rightarrow b,t, C^{(8)}, C^{(10)} ) \right\}.
\end{eqnarray*}
\par
Then we can obtain the form factors in the renormalized amplitude of the
t-channel vertex diagrams in the subprocess $\gamma\gamma \rightarrow
\tilde{\chi}^{+}_1 \bar{\tilde{\chi}}^{+}_1$.
\begin{eqnarray*}
f_{i}^{v,\hat{t}}=0~~(i=2,3,6,7,9,10,12,13,16 \sim 22),
\end{eqnarray*}
\begin{eqnarray*}
f_1^{v,\hat{t}}&=&2 i e (p_1 \cdot p_3) A_{t} \left\{
  m_{\tilde{\chi}^{+}_1}\left[g_{17}(p_1,p_3-p_1) +g_{17}(p_1-p_3,p_2)\right]
  \right. \\
&-&\left. g_{18}(p_1-p_3,p_2)-g_{19}(p_1,p_3-p_1) \right\},
\end{eqnarray*}
\begin{eqnarray*}
f_4^{v,\hat{t}}=2 i e (p_1 \cdot p_3) A_{t} \left[
       g_{12}(p_1-p_3,p_2)-g_{11}(p_1-p_3,p_2) \right],
\end{eqnarray*}
\begin{eqnarray*}
f_5^{v,\hat{t}}&=&-2 i e A_{t} \left\{
      i e \left[ C^{+}+C^{-} \right] +m_{\tilde{\chi}^{+}_1}^2 (g_{11}(p_1,p_3-p_1)-
     g_{12}(p_1,p_3-p_1)\right. \\
&-&\left. g_{13}(p_1,p_3-p_1)+g_{14}(p_1,p_3-p_1)-g_{17}(p_1,p_3-p_1)+
     g_{17}(p_1-p_3,p_2)) \right. \\
&+&\left. (p_1 \cdot p_3) (g_{13}(p_1,p_3-p_1)- g_{14}(p_1,p_3-p_1)+
     2 g_{17}(p_1,p_3-p_1)) \right. \\
&+&\left. m_{\tilde{\chi}^{+}_1}(g_{15}(p_1,p_3-p_1)-g_{16}(p_1,p_3-p_1)
     +g_{18}(p_1,p_3-p_1)-g_{18}(p_1-p_3,p_2) \right. \\
&-&\left. g_{19}(p_1,p_3-p_1)-
     g_{19}(p_1-p_3,p_2) )+g_{20}(p_1,p_3-p_1)+g_{20}(p_1-p_3,p_2) \right\},
\end{eqnarray*}
\begin{eqnarray*}
f_8^{v,\hat{t}}&=& 2 f_{14}^{v,\hat{t}} =2 i e A_{t} \left\{
     m_{\tilde{\chi}^{+}_1} \left[ g_{11}(p_1-p_3,p_2)-g_{12}(p_1-p_3,p_2)-
     g_{13}(p_1-p_3,p_2) \right.\right.\\
&+&\left. \left. g_{14}(p_1-p_3,p_2)-2 g_{17}(p_1-p_3,p_2)\right]+
     g_{15}(p_1-p_3,p_2)- g_{16}(p_1-p_3,p_2) \right. \\
&+&\left. 2 g_{18}(p_1-p_3,p_2) \right\},
\end{eqnarray*}
\begin{eqnarray*}
f_{11}^{v,\hat{t}}&=&-i e A_{t} \left\{
     i e (C^{+}+C^{-})+m_{\tilde{\chi}^{+}_1}^2(g_{17}(p_1,p_3-p_1)+
     g_{17}(p_1-p_3,p_2)) \right. \\
&-& \left. m_{\tilde{\chi}^{+}_1}(g_{18}(p_1,p_3-p_1)+g_{18}(p_1-p_3,p_2)
     +g_{19}(p_1,p_3-p_1) \right. \\
&+& \left. g_{19}(p_1-p_3,p_2))+g_{20}(p_1,p_3-p_1)+g_{20}(p_1-p_3,p_2)\right\},
\end{eqnarray*}
\begin{eqnarray*}
f_{15}^{v,\hat{t}}=
     f_{14}^{v,\hat{t}}(g_{i}(p_1-p_3,p_2) \rightarrow g_{i}(p_1,p_3-p_1)).
\end{eqnarray*}
\par
The form factors from the renormalized amplitude of t-channel box
diagrams Fig.1(c) are written as:
\begin{eqnarray*}
f_1^{b,\hat{t}} &=&
\left\{ \frac{-i e^2  Q_b^2}{32 \pi^2} \sum_{k=1,2}
    F_{1}^{b\tilde{t}_{k}\tilde{\chi}^{+}_{1}} m_{\tilde{\chi}_1^{+}} \left[ 2
    p_1 \cdot p_2 (D^{(1,k)}_{13}+D^{(1,k)}_{35}+2 D^{(1,k)}_{25}-D^{(1,k)}_{23}-
    D^{(1,k)}_{37}) \right. \right.\\
&& +2 p_1 \cdot p_3 (D^{(1,k)}_{11}+D^{(1,k)}_{12}+D^{(1,k)}_{21}+D^{(1,k)}_{23}+
    D^{(1,k)}_{34}+D^{(1,k)}_{37}+2 D^{(1,k)}_{24}-3 D^{(1,k)}_{25}\\
&& -D^{(1,k)}_{26}-D^{(1,k)}_{310}-D^{(1,k)}_{35}-2 D^{(1,k)}_{13})+2 p_2 \cdot p_3 (
    D^{(1,k)}_{23}+D^{(1,k)}_{37}-D^{(1,k)}_{13}-D^{(1,k)}_{25}\\
&& -D^{(1,k)}_{26}-D^{(1,k)}_{310})+m_{\tilde{\chi}_1^{+}}^2 (2 D^{(1,k)}_{13}+2
    D^{(1,k)}_{35}+4 D^{(1,k)}_{25}-3 D^{(1,k)}_{11}-3 D^{(1,k)}_{21}-2 D^{(1,k)}_{23}\\
&& \left.\left.-2 D^{(1,k)}_{37}-D^{(1,k)}_{0}-D^{(1,k)}_{31})+m_b^2 (D^{(1,k)}_{0}+
    D^{(1,k)}_{11})+4 D^{(1,k)}_{27}+(4-\epsilon) D^{(1,k)}_{311} \right.\right]\\
&& \left.+ F_{2}^{b\tilde{t}_{k}\tilde{\chi}^{+}_{1}} m_b \left[ 2 p_1 \cdot p_2 (
    D^{(1,k)}_{13}+D^{(1,k)}_{25}-D^{(1,k)}_{23})+2 p_1 \cdot p_3 (D^{(1,k)}_{11}+
    D^{(1,k)}_{12} \right. \right.\\
&& +D^{(1,k)}_{23}+D^{(1,k)}_{24}-D^{(1,k)}_{25}-D^{(1,k)}_{26}-2 D^{(1,k)}_{13})+2
    p_2 \cdot p_3 (D^{(1,k)}_{23}-D^{(1,k)}_{13}-D^{(1,k)}_{26})\\
&&  \left. \left. +m_{\tilde{\chi}_1^{+}}^2 (2 D^{(1,k)}_{13}+2 D^{(1,k)}_{25}-
    2 D^{(1,k)}_{11}-2 D^{(1,k)}_{23}-D^{(1,k)}_{0}-D^{(1,k)}_{21})
   + m_b^2 D^{(1,k)}_{0}+2 D^{(1,k)}_{27} \right.\right] \\
&& \left.+ (b,t,F_{1,2},D^{(1,k)} \rightarrow t,b,-F_{1,2},D^{(2,k)}) \right\} \\
&-& \left\{ \frac{i e^2 Q_t^2}{16 \pi^2} \sum_{k=1,2}
    (F_{1}^{b\tilde{t}_{k}\tilde{\chi}^{+}_{1}} m_{\tilde{\chi}_1^{+}} D^{(3,k)}_{311}
    -F_{2}^{b\tilde{t}_{k}\tilde{\chi}^{+}_{1}} m_b D^{(3,k)}_{27})
    + (b,t,F_{1,2},D^{(3,k)} \rightarrow t,b,-F_{1,2},D^{(4,k)}) \right\} \\
&-& \left\{ \frac{i e^2 Q_b Q_t}{16 \pi^2} \sum_{k=1,2}
    F_{1}^{b\tilde{t}_{k}\tilde{\chi}^{+}_{1}} m_{\tilde{\chi}_1^{+}} (
    D^{(5,k)}_{311}-D^{(5,k)}_{313})-F_{2}^{b\tilde{t}_{k}\tilde{\chi}^{+}_{1}} m_b
    D^{(5,k)}_{27} \right. \\
&& \left.+(b,t,F_{1,2},D^{(5,k)} \rightarrow t,b,-F_{1,2},D^{(6,k)}) \right\},
\end{eqnarray*}
\begin{eqnarray*}
f_2^{b,\hat{t}} &=&
\left\{ \frac{i e^2 Q_b^2}{16 \pi^2} \sum_{k=1,2}
    F_{1}^{b\tilde{t}_{k}\tilde{\chi}^{+}_{1}} m_{\tilde{\chi}_1^{+}} (D^{(1,k)}_{27}
    +D^{(1,k)}_{311})+F_{2}^{b\tilde{t}_{k}\tilde{\chi}^{+}_{1}} m_b D^{(1,k)}_{27}
    \right.\\
&&\left.+ (b,t,F_{1,2},D^{(1,k)} \rightarrow t,b,-F_{1,2},D^{(2,k)}) \right\} \\
&-& \left\{ \frac{i e^2 Q_t^2}{16 \pi^2} \sum_{k=1,2}
   ( F_{1}^{b\tilde{t}_{k}\tilde{\chi}^{+}_{1}} m_{\tilde{\chi}_1^{+}} D^{(3,k)}_{311}
    -F_{2}^{b\tilde{t}_{k}\tilde{\chi}^{+}_{1}} m_b D^{(3,k)}_{27})
    + (b,t,F_{1,2},D^{(3,k)} \rightarrow t,b,-F_{1,2},D^{(4,k)}) \right\} \\
&-& \left\{ \frac{i e^2 Q_b Q_t}{16 \pi^2} \sum_{k=1,2}
    F_{1}^{b\tilde{t}_{k}\tilde{\chi}^{+}_{1}} m_{\tilde{\chi}_1^{+}} (
    D^{(5,k)}_{311}-D^{(5,k)}_{313})-F_{2}^{b\tilde{t}_{k}\tilde{\chi}^{+}_{1}} m_b
    D^{(5,k)}_{27} \right. \\
&&\left.+ (b,t,F_{1,2},D^{(5,k)} \rightarrow t,b,-F_{1,2},D^{(6,k)}) \right\},
\end{eqnarray*}
\begin{eqnarray*}
f_3^{b,\hat{t}} &=&
\left\{ \frac{-i e^2 Q_b^2}{16 \pi^2} \sum_{k=1,2}
    F_{1}^{b\tilde{t}_{k}\tilde{\chi}^{+}_{1}} \left[ 2 p_1 \cdot p_2 (D^{(1,k)}_{25}
    +D^{(1,k)}_{35}+D^{(1,k)}_{39}-D^{(1,k)}_{26}-D^{(1,k)}_{310}-D^{(1,k)}_{37}) \right.
    \right.\\
&& +2 p_1 \cdot p_3 (D^{(1,k)}_{24}+D^{(1,k)}_{26}+D^{(1,k)}_{34}+D^{(1,k)}_{37}+
    D^{(1,k)}_{38}-D^{(1,k)}_{22}-D^{(1,k)}_{25}-D^{(1,k)}_{35}\\
&& -D^{(1,k)}_{36}-D^{(1,k)}_{39})+2 p_2 \cdot p_3 (D^{(1,k)}_{26}+D^{(1,k)}_{37}+
    D^{(1,k)}_{38}-D^{(1,k)}_{25}-D^{(1,k)}_{310}-D^{(1,k)}_{39})\\
&& +m_{\tilde{\chi}_1^{+}}^2 (D^{(1,k)}_{12}+D^{(1,k)}_{34}+2 D^{(1,k)}_{24}+2
    D^{(1,k)}_{25}+2 D^{(1,k)}_{35}+2 D^{(1,k)}_{39}-D^{(1,k)}_{11}-D^{(1,k)}_{31}\\
&& \left. \left. -2 D^{(1,k)}_{21}-2 D^{(1,k)}_{26}-2 D^{(1,k)}_{310}-2 D^{(1,k)}_{37})+
    m_b^2( D^{(1,k)}_{11}-D^{(1,k)}_{12}) + (4-\epsilon) (D^{(1,k)}_{311}-D^{(1,k)}_{312})
    \right.\right] \\
&& \left. +(b,t,F_{1,2},D^{(1,k)} \rightarrow t,b,-F_{1,2},D^{(2,k)}) \right\} \\
&-& \left\{ \frac{i e^2 Q_t^2}{8 \pi^2} \sum_{k=1,2}
    F_{1}^{b\tilde{t}_{k}\tilde{\chi}^{+}_{1}} (D^{(3,k)}_{311}-D^{(3,k)}_{312})
   + (b,t,F_{1,2},D^{(3,k)} \rightarrow t,b,-F_{1,2},D^{(4,k)}) \right\} \\
&-& \left\{ \frac{i e^2 Q_b Q_t}{8 \pi^2} \sum_{k=1,2}
  F_{1}^{b\tilde{t}_{k}\tilde{\chi}^{+}_{1}} (D^{(5,k)}_{311}-D^{(5,k)}_{312})
  + (b,t,F_{1,2},D^{(5,k)} \rightarrow t,b,-F_{1,2},D^{(6,k)}) \right\},
\end{eqnarray*}
\begin{eqnarray*}
f_4^{b,\hat{t}} &=&
\left\{ \frac{-i e^2 Q_b^2}{16 \pi^2} \sum_{k=1,2}
    F_{1}^{b\tilde{t}_{k}\tilde{\chi}^{+}_{1}} \left[ 2 p_1 \cdot p_2 (D^{(1,k)}_{23}
    +D^{(1,k)}_{37}+D^{(1,k)}_{39}-D^{(1,k)}_{26}-D^{(1,k)}_{310}-D^{(1,k)}_{33}) \right.
    \right.\\
&& +2 p_1 \cdot p_3 (2 D^{(1,k)}_{26}+2 D^{(1,k)}_{310}+D^{(1,k)}_{25}+D^{(1,k)}_{33}+
    D^{(1,k)}_{38}-2 D^{(1,k)}_{23}-2 D^{(1,k)}_{39}-D^{(1,k)}_{22}\\
&& -D^{(1,k)}_{36}-D^{(1,k)}_{37})+2 p_2 \cdot p_3 (D^{(1,k)}_{26}+D^{(1,k)}_{33}+
    D^{(1,k)}_{38}-2 D^{(1,k)}_{39}-D^{(1,k)}_{23})\\
&& +m_{\tilde{\chi}_1^{+}}^2 (D^{(1,k)}_{12}+D^{(1,k)}_{34}+2 D^{(1,k)}_{23}+2
    D^{(1,k)}_{24}+2 D^{(1,k)}_{37}+2 D^{(1,k)}_{39}-D^{(1,k)}_{13}-D^{(1,k)}_{35}\\
&& -2 D^{(1,k)}_{25}-2 D^{(1,k)}_{26}-2 D^{(1,k)}_{310}-2 D^{(1,k)}_{33})+m_b^2 (
    D^{(1,k)}_{13}-D^{(1,k)}_{12})\\
&& \left. \left.+(6-\epsilon) D^{(1,k)}_{313}-2 D^{(1,k)}_{27}
    -(4-\epsilon) D^{(1,k)}_{312} \right]
    + (b,t,F_{1,2},D^{(1,k)} \rightarrow t,b,-F_{1,2},D^{(2,k)}) \right\} \\
&+& \left\{ \frac{i e^2 Q_t^2}{8 \pi^2} \sum_{k=1,2}
   F_{1}^{b\tilde{t}_{k}\tilde{\chi}^{+}_{1}} (D^{(3,k)}_{27}+D^{(3,k)}_{312})
   + (b,t,F_{1,2},D^{(3,k)} \rightarrow t,b,-F_{1,2},D^{(4,k)}) \right\} \\
&-& \left\{ \frac{i e^2 Q_b Q_t}{8 \pi^2} \sum_{k=1,2}
    F_{1}^{b\tilde{t}_{k}\tilde{\chi}^{+}_{1}} (D^{(5,k)}_{313}-D^{(5,k)}_{27}-
    D^{(5,k)}_{312})
    + (b,t,F_{1,2},D^{(5,k)} \rightarrow t,b,-F_{1,2},D^{(6,k)}) \right\},
\end{eqnarray*}
\begin{eqnarray*}
f_5^{b,\hat{t}} &=&
\left\{ \frac{-i e^2 Q_b^2}{16 \pi^2} \sum_{k=1,2}
    F_{1}^{b\tilde{t}_{k}\tilde{\chi}^{+}_{1}} \left[ 2 p_2 \cdot p_3 (D^{(1,k)}_{25}
    -D^{(1,k)}_{26})+m_{\tilde{\chi}_1^{+}}^2 (D^{(1,k)}_{0}+D^{(1,k)}_{21}+2
    D^{(1,k)}_{11} \right. \right.\\
&& \left.- 2 D^{(1,k)}_{13}-2 D^{(1,k)}_{25})+m_b^2 D^{(1,k)}_{0}+2 (D^{(1,k)}_{313}-
    D^{(1,k)}_{311}) \right]+2 F_{2}^{b\tilde{t}_{k}\tilde{\chi}^{+}_{1}} m_b
    m_{\tilde{\chi}_1^{+}} \cdot \\
&&  (D^{(1,k)}_{0}+D^{(1,k)}_{11}-D^{(1,k)}_{13})
    + \left. (b,t,F_{1,2},D^{(1,k)} \rightarrow t,b,-F_{1,2},D^{(2,k)}) \right\}\\
&-& \left\{ \frac{i e^2 Q_t^2}{8 \pi^2} \sum_{k=1,2}
    F_{1}^{b\tilde{t}_{k}\tilde{\chi}^{+}_{1}} (D^{(3,k)}_{27}+D^{(3,k)}_{311}-
    D^{(3,k)}_{313})
    + (b,t,F_{1,2},D^{(3,k)} \rightarrow t,b,-F_{1,2},D^{(4,k)}) \right\} \\
&-& \left\{ \frac{i e^2 Q_b Q_t}{16 \pi^2} \sum_{k=1,2}
    F_{1}^{b\tilde{t}_{k}\tilde{\chi}^{+}_{1}} \left[ 2 p_1 \cdot p_2 (2
    D^{(5,k)}_{25}+D^{(5,k)}_{13}+D^{(5,k)}_{35})-2 p_1 \cdot p_3 (2 D^{(5,k)}_{24}+
    D^{(5,k)}_{12} \right. \right.\\
&& +D^{(5,k)}_{34})-2 p_2 \cdot p_3 (D^{(5,k)}_{13}+D^{(5,k)}_{25}+D^{(5,k)}_{26}+
    D^{(5,k)}_{310})-m_b^2 (D^{(5,k)}_{0}+D^{(5,k)}_{11})\\
&& \left. \left.+ m_{\tilde{\chi}_1^{+}}^2 (3 D^{(5,k)}_{21}+D^{(5,k)}_{23}+D^{(5,k)}_{31}+
    D^{(5,k)}_{37}+2 D^{(5,k)}_{11})-4 D^{(5,k)}_{27}-(4-\epsilon) D^{(5,k)}_{311} \right. \right]\\
&& \left. -F_{2}^{b\tilde{t}_{k}\tilde{\chi}^{+}_{1}} 2 m_b m_{\tilde{\chi}^{+}_1}
    (D^{(5,k)}_{0}+D^{(5,k)}_{11})
    + (b,t,F_{1,2},D^{(5,k)} \rightarrow t,b,-F_{1,2},D^{(6,k)}) \right\},
\end{eqnarray*}
\begin{eqnarray*}
f_6^{b,\hat{t}} &=&
\left\{ \frac{-i e^2 Q_b^2}{16 \pi^2} \sum_{k=1,2}
    F_{1}^{b\tilde{t}_{k}\tilde{\chi}^{+}_{1}} \left[ 2 p_1 \cdot p_2 (D^{(1,k)}_{33}
    -D^{(1,k)}_{37})+2 p_1 \cdot p_3 (D^{(1,k)}_{23}+D^{(1,k)}_{37}+D^{(1,k)}_{39}
     \right. \right.\\
&& -D^{(1,k)}_{25}-D^{(1,k)}_{310}-D^{(1,k)}_{33})+2 p_2 \cdot p_3 (D^{(1,k)}_{39}-
    D^{(1,k)}_{33})+m_{\tilde{\chi}_1^{+}}^2 (D^{(1,k)}_{35}+2 D^{(1,k)}_{33}\\
&& \left. \left. -D^{(1,k)}_{13}-2 D^{(1,k)}_{37})-m_b^2 D^{(1,k)}_{13}-(4-\epsilon)
    D^{(1,k)}_{313} \right]-2 F_{2}^{b\tilde{t}_{k}\tilde{\chi}^{+}_{1}} m_b
    m_{\tilde{\chi}_1^{+}} D^{(1,k)}_{13} \right. \\
&& \left.+(b,t,F_{1,2},D^{(1,k)} \rightarrow t,b,-F_{1,2},D^{(2,k)}) \right\} \\
&+& \left\{ \frac{i e^2 Q_t^2}{8 \pi^2} \sum_{k=1,2}
     F_{1}^{b\tilde{t}_{k}\tilde{\chi}^{+}_{1}} D^{(3,k)}_{313}
    + (b,t,F_{1,2},D^{(3,k)} \rightarrow t,b,-F_{1,2},D^{(4,k)}) \right\} \\
&-& \left\{ \frac{i e^2 Q_b Q_t}{16 \pi^2} \sum_{k=1,2}
    F_{1}^{b\tilde{t}_{k}\tilde{\chi}^{+}_{1}} \left[ 2 p_1 \cdot p_2 (D^{(5,k)}_{23}
    +D^{(5,k)}_{37})-2 p_1 \cdot p_3 (D^{(5,k)}_{26}+D^{(5,k)}_{310}) \right. \right.
    \\
&& -2 p_2 \cdot p_3 (D^{(5,k)}_{23}+D^{(5,k)}_{39})-m_b^2 D^{(5,k)}_{13}+
    m_{\tilde{\chi}_1^{+}}^2 (D^{(5,k)}_{33}+D^{(5,k)}_{35}+2 D^{(5,k)}_{25})\\
&& \left. \left.-(4-\epsilon) D^{(5,k)}_{313} \right]-
    2 F_{2}^{b\tilde{t}_{k}\tilde{\chi}^{+}_{1}} m_b m_{\tilde{\chi}_1^{+}}
    D^{(5,k)}_{13}
    + (b,t,F_{1,2},D^{(5,k)} \rightarrow t,b,-F_{1,2},D^{(6,k)}) \right\},
\end{eqnarray*}
\begin{eqnarray*}
f_7^{b,\hat{t}} &=&
\left\{ \frac{-i e^2 Q_b^2}{8 \pi^2} \sum_{k=1,2}
    F_{1}^{b\tilde{t}_{k}\tilde{\chi}^{+}_{1}} m_{\tilde{\chi}_1^{+}} (D^{(1,k)}_{11}
    -D^{(1,k)}_{12}+2 D^{(1,k)}_{21}-2 D^{(1,k)}_{24}-D^{(1,k)}_{25}+D^{(1,k)}_{26}
     \right.\\
&& +D^{(1,k)}_{31}+D^{(1,k)}_{310}-D^{(1,k)}_{34}-D^{(1,k)}_{35})+
    F_{2}^{b\tilde{t}_{k}\tilde{\chi}^{+}_{1}} m_b (D^{(1,k)}_{11}-D^{(1,k)}_{12}+
    D^{(1,k)}_{21}\\
&& \left. -D^{(1,k)}_{24}-D^{(1,k)}_{25}+D^{(1,k)}_{26})
   + (b,t,F_{1,2},D^{(1,k)} \rightarrow t,b,-F_{1,2},D^{(2,k)}) \right\} \\
&-& \left\{ \frac{i e^2 Q_t^2}{8 \pi^2} \sum_{k=1,2}
  F_{1}^{b\tilde{t}_{k}\tilde{\chi}^{+}_{1}} m_{\tilde{\chi}_1^{+}} (D^{(3,k)}_{24}
  +D^{(3,k)}_{34}+D^{(3,k)}_{35}-D^{(3,k)}_{21}-D^{(3,k)}_{310}-D^{(3,k)}_{31}) \right.\\
&& +F_{2}^{b\tilde{t}_{k}\tilde{\chi}^{+}_{1}} m_b (D^{(3,k)}_{11}+
    D^{(3,k)}_{21}+D^{(3,k)}_{26}-D^{(3,k)}_{12}-D^{(3,k)}_{24}-D^{(3,k)}_{25})  \\
&& \left. +(b,t,F_{1,2},D^{(3,k)} \rightarrow t,b,-F_{1,2},D^{(4,k)}) \right\} \\
&-& \left\{ \frac{i e^2 Q_b Q_t}{8 \pi^2} \sum_{k=1,2}
    F_{1}^{b\tilde{t}_{k}\tilde{\chi}^{+}_{1}} m_{\tilde{\chi}_1^{+}} (D^{(5,k)}_{24}
    +D^{(5,k)}_{25}+D^{(5,k)}_{34}+D^{(5,k)}_{35}-D^{(5,k)}_{21}-D^{(5,k)}_{26} \right.\\
&& -D^{(5,k)}_{310}-D^{(5,k)}_{31})+F_{2}^{b\tilde{t}_{k}\tilde{\chi}^{+}_{1}}
    m_b (D^{(5,k)}_{11}+D^{(5,k)}_{21}-D^{(5,k)}_{12}-D^{(5,k)}_{24}) \\
&& \left.+(b,t,F_{1,2},D^{(5,k)} \rightarrow t,b,-F_{1,2},D^{(6,k)}) \right\},
\end{eqnarray*}
\begin{eqnarray*}
f_8^{b,\hat{t}} &=&
\left\{ \frac{-i e^2 Q_b^2}{8 \pi^2} \sum_{k=1,2}
    F_{1}^{b\tilde{t}_{k}\tilde{\chi}^{+}_{1}} m_{\tilde{\chi}_1^{+}} (D^{(1,k)}_{26}
    +D^{(1,k)}_{310}-2 D^{(1,k)}_{24}-D^{(1,k)}_{12}-D^{(1,k)}_{34})\right. \\
&&  + F_{2}^{b\tilde{t}_{k}\tilde{\chi}^{+}_{1}} m_b (D^{(1,k)}_{26}-
    D^{(1,k)}_{12}-D^{(1,k)}_{24}) \\
&& \left. +(b,t,F_{1,2},D^{(1,k)} \rightarrow t,b,-F_{1,2},D^{(2,k)}) \right\} \\
&-& \left\{ \frac{i e^2 Q_t^2}{8 \pi^2} \sum_{k=1,2}
  F_{1}^{b\tilde{t}_{k}\tilde{\chi}^{+}_{1}} m_{\tilde{\chi}_1^{+}} (D^{(3,k)}_{11}
 +D^{(3,k)}_{21}+D^{(3,k)}_{24}+D^{(3,k)}_{34}-D^{(3,k)}_{25}-D^{(3,k)}_{310}) \right. \\
&& +F_{2}^{b\tilde{t}_{k}\tilde{\chi}^{+}_{1}} m_b (-D^{(3,k)}_{0}+
    D^{(3,k)}_{13}+D^{(3,k)}_{26}-D^{(3,k)}_{11}-D^{(3,k)}_{12}-D^{(3,k)}_{24})  \\
&& \left.+(b,t,F_{1,2},D^{(3,k)} \rightarrow t,b,-F_{1,2},D^{(4,k)}) \right\} \\
&-& \left\{ \frac{i e^2 Q_b Q_t}{8 \pi^2} \sum_{k=1,2}
    F_{1}^{b\tilde{t}_{k}\tilde{\chi}^{+}_{1}} m_{\tilde{\chi}_1^{+}} (D^{(5,k)}_{11}
    +D^{(5,k)}_{21}+D^{(5,k)}_{23}+D^{(5,k)}_{24}+D^{(5,k)}_{34}+D^{(5,k)}_{37} \right.\\
&& -2 D^{(5,k)}_{25}-D^{(5,k)}_{13}-D^{(5,k)}_{26}-D^{(5,k)}_{310}-D^{(5,k)}_{35})+
    F_{2}^{b\tilde{t}_{k}\tilde{\chi}^{+}_{1}} m_b (D^{(5,k)}_{13}+D^{(5,k)}_{25}\\
&& \left. -D^{(5,k)}_{0}-D^{(5,k)}_{11}-D^{(5,k)}_{12}-D^{(5,k)}_{24})
  + (b,t,F_{1,2},D^{(5,k)} \rightarrow t,b,-F_{1,2},D^{(6,k)}) \right\},
\end{eqnarray*}
\begin{eqnarray*}
f_9^{b,\hat{t}} &=&
\left\{ \frac{-i e^2 Q_b^2}{8 \pi^2} \sum_{k=1,2}
    F_{1}^{b\tilde{t}_{k}\tilde{\chi}^{+}_{1}} m_{\tilde{\chi}_1^{+}} (D^{(1,k)}_{26}
    +D^{(1,k)}_{310}-D^{(1,k)}_{25}-D^{(1,k)}_{35}) \right.\\
&& \left.+F_{2}^{b\tilde{t}_{k}\tilde{\chi}^{+}_{1}} m_b (D^{(1,k)}_{26}-
    D^{(1,k)}_{25})
  + (b,t,F_{1,2},D^{(1,k)} \rightarrow t,b,-F_{1,2},D^{(2,k)}) \right\} \\
&-& \left\{ \frac{i e^2 Q_t^2}{8 \pi^2} \sum_{k=1,2}
    F_{1}^{b\tilde{t}_{k}\tilde{\chi}^{+}_{1}} m_{\tilde{\chi}_1^{+}} (D^{(3,k)}_{35}
    -D^{(3,k)}_{310})+F_{2}^{b\tilde{t}_{k}\tilde{\chi}^{+}_{1}} m_b (D^{(3,k)}_{26}-
    D^{(3,k)}_{25}) \right. \\
&& \left.+(b,t,F_{1,2},D^{(3,k)} \rightarrow t,b,-F_{1,2},D^{(4,k)}) \right\} \\
&-& \left\{ \frac{i e^2 Q_b Q_t}{8 \pi^2} \sum_{k=1,2}
    F_{1}^{b\tilde{t}_{k}\tilde{\chi}^{+}_{1}} m_{\tilde{\chi}_1^{+}} (
    D^{(5,k)}_{310}+D^{(5,k)}_{37}-D^{(5,k)}_{35}-D^{(5,k)}_{39}) \right.\\
&& \left.+F_{2}^{b\tilde{t}_{k}\tilde{\chi}^{+}_{1}} m_b (D^{(5,k)}_{25}-
    D^{(5,k)}_{26})
    + (b,t,F_{1,2},D^{(5,k)} \rightarrow t,b,-F_{1,2},D^{(6,k)}) \right\},
\end{eqnarray*}
\begin{eqnarray*}
f_{10}^{b,\hat{t}} &=&
\left\{ \frac{-i e^2 Q_b^2}{8 \pi^2} \sum_{k=1,2}
    F_{1}^{b\tilde{t}_{k}\tilde{\chi}^{+}_{1}} m_{\tilde{\chi}_1^{+}} (D^{(1,k)}_{26}
    +D^{(1,k)}_{310})+F_{2}^{b\tilde{t}_{k}\tilde{\chi}^{+}_{1}} m_b D^{(1,k)}_{26}
    \right. \\
&& \left.+(b,t,F_{1,2},D^{(1,k)} \rightarrow t,b,-F_{1,2},D^{(2,k)}) \right\} \\
&-&\left\{ \frac{i e^2 Q_t^2}{8 \pi^2} \sum_{k=1,2} \left[ -
    F_{1}^{b\tilde{t}_{k}\tilde{\chi}^{+}_{1}} m_{\tilde{\chi}_1^{+}} (D^{(3,k)}_{25}
    +D^{(3,k)}_{310})+F_{2}^{b\tilde{t}_{k}\tilde{\chi}^{+}_{1}} m_b (D^{(3,k)}_{13}+
    D^{(3,k)}_{26}) \right.\right] \\
&& \left. +(b,t,F_{1,2},D^{(3,k)} \rightarrow t,b,-F_{1,2},D^{(4,k)}) \right\} \\
&-& \left\{ \frac{i e^2 Q_b Q_t}{8 \pi^2} \sum_{k=1,2}
    F_{1}^{b\tilde{t}_{k}\tilde{\chi}^{+}_{1}} m_{\tilde{\chi}_1^{+}} (D^{(5,k)}_{25}
    +D^{(5,k)}_{310}+D^{(5,k)}_{33}-D^{(5,k)}_{23}-D^{(5,k)}_{37}-D^{(5,k)}_{39}) \right.
    \\
&& \left.+F_{2}^{b\tilde{t}_{k}\tilde{\chi}^{+}_{1}} m_b (D^{(5,k)}_{23}-
    D^{(5,k)}_{13}-D^{(5,k)}_{26})
    + (b,t,F_{1,2},D^{(5,k)} \rightarrow t,b,-F_{1,2},D^{(6,k)}) \right\},
\end{eqnarray*}
\begin{eqnarray*}
f_{11}^{b,\hat{t}} &=&
\left\{ \frac{-i e^2 Q_b^2}{32 \pi^2} \sum_{k=1,2}
    F_{1}^{b\tilde{t}_{k}\tilde{\chi}^{+}_{1}} \left[ 2 p_1 \cdot p_2 (D^{(1,k)}_{25}
    -D^{(1,k)}_{26}-D^{(1,k)}_{310}-D^{(1,k)}_{33}+D^{(1,k)}_{37}+D^{(1,k)}_{39}) \right.
     \right.\\
&& +2 p_1 \cdot p_3 (-D^{(1,k)}_{22}-D^{(1,k)}_{23}+2 D^{(1,k)}_{26}+2 D^{(1,k)}_{310}+
    D^{(1,k)}_{33}-D^{(1,k)}_{36}-D^{(1,k)}_{37}+D^{(1,k)}_{38}\\
&& -2 D^{(1,k)}_{39})+2 p_2 \cdot p_3 (D^{(1,k)}_{33}+D^{(1,k)}_{38}-2 D^{(1,k)}_{39})+
    m_{\tilde{\chi}_1^{+}}^2 (D^{(1,k)}_{0}+D^{(1,k)}_{12}-D^{(1,k)}_{13}\\
&& -D^{(1,k)}_{21}+2 D^{(1,k)}_{24}-2 D^{(1,k)}_{26}-2 D^{(1,k)}_{310}-2 D^{(1,k)}_{33}+
    D^{(1,k)}_{34}-D^{(1,k)}_{35}+2 D^{(1,k)}_{37}+2 D^{(1,k)}_{39})\\
&& \left. \left. +m_b^2 (D^{(1,k)}_{0}-D^{(1,k)}_{12}+D^{(1,k)}_{13})+(4-\epsilon)
   (D^{(1,k)}_{313}-D^{(1,k)}_{312}) \right]+2 F_{2}^{b\tilde{t}_{k}\tilde{\chi}^{+}_{1}}
   m_b m_{\tilde{\chi}_1^{+}} D^{(1,k)}_{0} \right. \\
&& \left. +(b,t,F_{1,2},D^{(1,k)} \rightarrow t,b,-F_{1,2},D^{(2,k)}) \right\} \\
&-& \left\{ \frac{i e^2 Q_t^2}{16 \pi^2} \sum_{k=1,2}
    F_{1}^{b\tilde{t}_{k}\tilde{\chi}^{+}_{1}} (D^{(3,k)}_{313}-D^{(3,k)}_{312})
   + (b,t,F_{1,2},D^{(3,k)} \rightarrow t,b,-F_{1,2},D^{(4,k)}) \right\} \\
&+& \left\{ \frac{i e^2 Q_b Q_t}{16 \pi^2} \sum_{k=1,2}
  F_{1}^{b\tilde{t}_{k}\tilde{\chi}^{+}_{1}} (D^{(5,k)}_{27}+D^{(5,k)}_{312})
   + (b,t,F_{1,2},D^{(5,k)} \rightarrow t,b,-F_{1,2},D^{(6,k)}) \right\},
\end{eqnarray*}
\begin{eqnarray*}
f_{12}^{b,\hat{t}} &=&
\left\{ \frac{-i e^2 Q_b^2}{16 \pi^2} \sum_{k=1,2}
    F_{1}^{b\tilde{t}_{k}\tilde{\chi}^{+}_{1}} (D^{(1,k)}_{27}+D^{(1,k)}_{312}-
    D^{(1,k)}_{313})
    + (b,t,F_{1,2},D^{(1,k)} \rightarrow t,b,-F_{1,2},D^{(2,k)}) \right\} \\
&-& \left\{ \frac{i e^2 Q_t^2}{16 \pi^2} \sum_{k=1,2}
    F_{1}^{b\tilde{t}_{k}\tilde{\chi}^{+}_{1}} (D^{(3,k)}_{313}-D^{(3,k)}_{312})
    + (b,t,F_{1,2},D^{(3,k)} \rightarrow t,b,-F_{1,2},D^{(4,k)}) \right\} \\
&+& \left\{ \frac{i e^2 Q_b Q_t}{16 \pi^2} \sum_{k=1,2}
    F_{1}^{b\tilde{t}_{k}\tilde{\chi}^{+}_{1}} D^{(5,k)}_{312}
    + (b,t,F_{1,2},D^{(5,k)} \rightarrow t,b,-F_{1,2},D^{(6,k)}) \right\},
\end{eqnarray*}
\begin{eqnarray*}
f_{13}^{b,\hat{t}} &=&
\left\{ \frac{-i e^2 Q_b^2}{16 \pi^2} \sum_{k=1,2}
    F_{1}^{b\tilde{t}_{k}\tilde{\chi}^{+}_{1}} m_{\tilde{\chi}_1^{+}} (D^{(1,k)}_{11}
    +D^{(1,k)}_{21}-D^{(1,k)}_{12}-D^{(1,k)}_{24}) \right.\\
&& \left.+F_{2}^{b\tilde{t}_{k}\tilde{\chi}^{+}_{1}} m_b (D^{(1,k)}_{11}-
    D^{(1,k)}_{12}) + (b,t,F_{1,2},D^{(1,k)} \rightarrow t,b,-F_{1,2},D^{(2,k)})
    \right\},
\end{eqnarray*}
\begin{eqnarray*}
f_{14}^{b,\hat{t}} &=&
\left\{ \frac{i e^2 Q_b^2}{16 \pi^2} \sum_{k=1,2}
    F_{1}^{b\tilde{t}_{k}\tilde{\chi}^{+}_{1}} m_{\tilde{\chi}_1^{+}} (D^{(1,k)}_{12}
    +D^{(1,k)}_{24})+F_{2}^{b\tilde{t}_{k}\tilde{\chi}^{+}_{1}} m_b D^{(1,k)}_{12}
    \right. \\
&& \left.+ (b,t,F_{1,2},D^{(1,k)} \rightarrow t,b,-F_{1,2},D^{(2,k)}) \right\},
\end{eqnarray*}
\begin{eqnarray*}
f_{15}^{b,\hat{t}} &=&
\left\{ \frac{-i e^2 Q_b^2}{16 \pi^2} \sum_{k=1,2}
    F_{1}^{b\tilde{t}_{k}\tilde{\chi}^{+}_{1}} m_{\tilde{\chi}_1^{+}} (D^{(1,k)}_{13}
    +D^{(1,k)}_{25}-D^{(1,k)}_{11}-D^{(1,k)}_{21}) \right.\\
&& \left.+F_{2}^{b\tilde{t}_{k}\tilde{\chi}^{+}_{1}} m_b (D^{(1,k)}_{13}-
    D^{(1,k)}_{11})
    + (b,t,F_{1,2},D^{(1,k)} \rightarrow t,b,-F_{1,2},D^{(2,k)}) \right\} \\
&-& \left\{ \frac{i e^2 Q_b Q_t}{16 \pi^2} \sum_{k=1,2}
    F_{1}^{b\tilde{t}_{k}\tilde{\chi}^{+}_{1}} m_{\tilde{\chi}_1^{+}} (D^{(5,k)}_{13}
    +D^{(5,k)}_{25}-D^{(5,k)}_{11}-D^{(5,k)}_{21}) \right.\\
&& \left.+F_{2}^{b\tilde{t}_{k}\tilde{\chi}^{+}_{1}} m_b (D^{(5,k)}_{0}+D^{(5,k)}_{11})
    + (b,t,F_{1,2},D^{(5,k)} \rightarrow t,b,-F_{1,2},D^{(6,k)}) \right\},
\end{eqnarray*}
\begin{eqnarray*}
f_{16}^{b,\hat{t}} &=&
\left\{ \frac{-i e^2 Q_b^2}{16 \pi^2} \sum_{k=1,2}
    F_{1}^{b\tilde{t}_{k}\tilde{\chi}^{+}_{1}} m_{\tilde{\chi}_1^{+}} (D^{(1,k)}_{13}
    +D^{(1,k)}_{25})+F_{2}^{b\tilde{t}_{k}\tilde{\chi}^{+}_{1}} m_b D^{(1,k)}_{13}
    \right. \\
&& \left.+(b,t,F_{1,2},D^{(1,k)} \rightarrow t,b,-F_{1,2},D^{(2,k)}) \right\} \\
&-& \left\{ \frac{i e^2 Q_b Q_t}{16 \pi^2} \sum_{k=1,2}
    F_{1}^{b\tilde{t}_{k}\tilde{\chi}^{+}_{1}} m_{\tilde{\chi}_1^{+}} (D^{(5,k)}_{23}
    -D^{(5,k)}_{25})+F_{2}^{b\tilde{t}_{k}\tilde{\chi}^{+}_{1}} m_b D^{(5,k)}_{13}
    \right. \\
&&\left.+(b,t,F_{1,2},D^{(5,k)} \rightarrow t,b,-F_{1,2},D^{(6,k)}) \right\},
\end{eqnarray*}
\begin{eqnarray*}
f_{17}^{b,\hat{t}} &=&
\left\{ \frac{-i e^2 Q_b^2}{8 \pi^2} \sum_{k=1,2}
    F_{1}^{b\tilde{t}_{k}\tilde{\chi}^{+}_{1}} (D^{(1,k)}_{22}+D^{(1,k)}_{25}+
    D^{(1,k)}_{35}+D^{(1,k)}_{36}+D^{(1,k)}_{39}-D^{(1,k)}_{24}-D^{(1,k)}_{26} \right.\\
&& \left.-D^{(1,k)}_{34}-D^{(1,k)}_{37}-D^{(1,k)}_{38})
    + (b,t,F_{1,2},D^{(1,k)} \rightarrow t,b,-F_{1,2},D^{(2,k)}) \right\} \\
&-& \left\{ \frac{i e^2 Q_t^2}{8 \pi^2} \sum_{k=1,2}
    F_{1}^{b\tilde{t}_{k}\tilde{\chi}^{+}_{1}} (D^{(3,k)}_{24}+D^{(3,k)}_{26}+
    D^{(3,k)}_{34}+D^{(3,k)}_{37}+D^{(3,k)}_{38}-D^{(3,k)}_{22}-D^{(3,k)}_{25}\right. \\
&& \left.-D^{(3,k)}_{35}-D^{(3,k)}_{36}-D^{(3,k)}_{39})
   + (b,t,F_{1,2},D^{(3,k)} \rightarrow t,b,-F_{1,2},D^{(4,k)}) \right\} \\
&-& \left\{ \frac{i e^2 Q_b Q_t}{8 \pi^2} \sum_{k=1,2}
    F_{1}^{b\tilde{t}_{k}\tilde{\chi}^{+}_{1}} (D^{(5,k)}_{24}+D^{(5,k)}_{34}-
    D^{(5,k)}_{22}-D^{(5,k)}_{36}) \right.\\
&&\left. + (b,t,F_{1,2},D^{(5,k)} \rightarrow t,b,-F_{1,2},D^{(6,k)}) \right\},
\end{eqnarray*}
\begin{eqnarray*}
f_{18}^{b,\hat{t}} &=&
\left\{ \frac{-i e^2 Q_b^2}{8 \pi^2} \sum_{k=1,2}
    F_{1}^{b\tilde{t}_{k}\tilde{\chi}^{+}_{1}} (D^{(1,k)}_{22}+D^{(1,k)}_{23}+
    D^{(1,k)}_{36}+D^{(1,k)}_{39}-D^{(1,k)}_{25}-D^{(1,k)}_{26}-D^{(1,k)}_{310}
    - D^{(1,k)}_{38}) \right. \\
&& \left. +(b,t,F_{1,2},D^{(1,k)} \rightarrow t,b,-F_{1,2},D^{(2,k)}) \right\} \\
&-& \left\{ \frac{i e^2 Q_t^2}{8 \pi^2} \sum_{k=1,2}
    F_{1}^{b\tilde{t}_{k}\tilde{\chi}^{+}_{1}} (2 D^{(3,k)}_{26}+D^{(3,k)}_{13}+
    D^{(3,k)}_{25}+D^{(3,k)}_{310}+D^{(3,k)}_{38}-D^{(3,k)}_{12}-D^{(3,k)}_{22}\right. \\
&& \left.-D^{(3,k)}_{23}-D^{(3,k)}_{24}-D^{(3,k)}_{36}-D^{(3,k)}_{39})
   + (b,t,F_{1,2},D^{(3,k)} \rightarrow t,b,-F_{1,2},D^{(4,k)}) \right\} \\
&-& \left\{ \frac{i e^2 Q_b Q_t}{8 \pi^2} \sum_{k=1,2}
    F_{1}^{b\tilde{t}_{k}\tilde{\chi}^{+}_{1}} (D^{(5,k)}_{13}+D^{(5,k)}_{25}+
    D^{(5,k)}_{26}+D^{(5,k)}_{310}-D^{(5,k)}_{12}-D^{(5,k)}_{22}-D^{(5,k)}_{24}-
    D^{(5,k)}_{36}) \right.\\
&& \left.+(b,t,F_{1,2},D^{(5,k)} \rightarrow t,b,-F_{1,2},D^{(6,k)}) \right\},
\end{eqnarray*}
\begin{eqnarray*}
f_{19}^{b,\hat{t}} &=&
\left\{ \frac{-i e^2 Q_b^2}{8 \pi^2} \sum_{k=1,2}
    F_{1}^{b\tilde{t}_{k}\tilde{\chi}^{+}_{1}} (D^{(1,k)}_{25}+D^{(1,k)}_{310}+
    D^{(1,k)}_{39}-D^{(1,k)}_{26}-D^{(1,k)}_{37}-D^{(1,k)}_{38}) \right. \\
&& \left. +(b,t,F_{1,2},D^{(1,k)} \rightarrow t,b,-F_{1,2},D^{(2,k)}) \right\} \\
&-& \left\{ \frac{i e^2 Q_t^2}{8 \pi^2} \sum_{k=1,2}
    F_{1}^{b\tilde{t}_{k}\tilde{\chi}^{+}_{1}} (D^{(3,k)}_{37}+D^{(3,k)}_{38}-
    D^{(3,k)}_{310}-D^{(3,k)}_{39}) \right. \\
&& \left. + (b,t,F_{1,2},D^{(3,k)} \rightarrow t,b,-F_{1,2},D^{(4,k)}) \right\} \\
&-& \left\{ \frac{i e^2 Q_b Q_t}{8 \pi^2} \sum_{k=1,2}
    F_{1}^{b\tilde{t}_{k}\tilde{\chi}^{+}_{1}} (D^{(5,k)}_{310}-D^{(5,k)}_{38})
    + (b,t,F_{1,2},D^{(5,k)} \rightarrow t,b,-F_{1,2},D^{(6,k)}) \right\},
\end{eqnarray*}
\begin{eqnarray*}
f_{20}^{b,\hat{t}} &=&
\left\{ \frac{-i e^2 Q_b^2}{8 \pi^2} \sum_{k=1,2}
    F_{1}^{b\tilde{t}_{k}\tilde{\chi}^{+}_{1}} (D^{(1,k)}_{23}+D^{(1,k)}_{39}-
    D^{(1,k)}_{26}-D^{(1,k)}_{38}) \right. \\
&&\left. + (b,t,F_{1,2},D^{(1,k)} \rightarrow t,b,-F_{1,2},D^{(2,k)}) \right\} \\
&-& \left\{ \frac{i e^2 Q_t^2}{8 \pi^2} \sum_{k=1,2}
    F_{1}^{b\tilde{t}_{k}\tilde{\chi}^{+}_{1}} (D^{(3,k)}_{26}+D^{(3,k)}_{38}-
    D^{(3,k)}_{23}-D^{(3,k)}_{39}) \right. \\
&&\left. + (b,t,F_{1,2},D^{(3,k)} \rightarrow t,b,-F_{1,2},D^{(4,k)}) \right\} \\
&-& \left\{ \frac{i e^2 Q_b Q_t}{8 \pi^2} \sum_{k=1,2}
    F_{1}^{b\tilde{t}_{k}\tilde{\chi}^{+}_{1}} (D^{(5,k)}_{23}+D^{(5,k)}_{39}-
    D^{(5,k)}_{26}-D^{(5,k)}_{38}) \right. \\
&&\left. + (b,t,F_{1,2},D^{(5,k)} \rightarrow t,b,-F_{1,2},D^{(6,k)}) \right\},
\end{eqnarray*}
\begin{eqnarray*}
f_{i}^{v,\hat{t}}=0~~(i=21\sim 22).
\end{eqnarray*}
\par
The form factors in the renormalized amplitude of the quartic interaction
diagrams Fig.1(d) are expressed as:
\begin{eqnarray*}
f_1^{q} &=& f_2^{q} = \left\{
\frac{i e^2 Q_t^2}{32 \pi^2} \sum_{k=1,2} (m_{\tilde{\chi}_1^{+}} C^{(1,k)}_{11}
    F_{1}^{b\tilde{t}_{k}\tilde{\chi}^{+}_{1}}-m_b C^{(1,k)}_{0}
    F_{2}^{b\tilde{t}_{k}\tilde{\chi}^{+}_{1}}) \right.\\
&+& \frac{e^2 Q_t^2}{16 \pi^2} \sum_{k=1,2} \bar{B}^{(1,k)}_{0}\left[
    (V_{h^0\tilde{t}_{k}\tilde{t}_{k}}
     V_{h^0\tilde{\chi}^{+}_{1}\tilde{\chi}^{+}_{1}}^{s}) A_h+(
    V_{H^0\tilde{t}_{k}\tilde{t}_{k}}
    V_{H^0\tilde{\chi}^{+}_{1}\tilde{\chi}^{+}_{1}}^{s}) A_H \right] \\
&& \left.+(t,b, F_{1,2},\bar{B}^{(1,k)},C^{(1,k)} \rightarrow
           b,t,-F_{1,2},\bar{B}^{(2,k)},C^{(2,k)}) \right\},
\end{eqnarray*}
$$
f_i^{q}=0,~(i=3 \sim 22)
$$
\par
The form factors in the renormalized amplitude from the t-channel triangle
diagrams depicted in Fig.1(e), are listed below:
\begin{eqnarray*}
f_1^{tr,\hat{t}} &=& f_2^{tr,\hat{t}} = \left\{
\frac{-e^2 Q_t^2}{8 \pi^2} \sum_{k=1,2} \left[ \bar{C}^{(3,k)}_{24} (
    V_{H^0\tilde{t}_{k}\tilde{t}_{k}}
    V_{H^0\tilde{\chi}^{+}_{1}\tilde{\chi}^{+}_{1}}^{s} A_{H}+
    V_{h^0\tilde{t}_{k}\tilde{t}_{k}}
    V_{h^0\tilde{\chi}^{+}_{1}\tilde{\chi}^{+}_{1}}^{s} A_{h}) \right]\right. \\
&+& \left. \frac{-e^2 Q_b^2}{8 \pi^2}
  (V_{H^0bb} V_{H^0\tilde{\chi}^{+}_{1}\tilde{\chi}^{+}_{1}}^{s} A_{H}+V_{h^0bb}
    V_{h^0\tilde{\chi}^{+}_{1}\tilde{\chi}^{+}_{1}}^{s} A_{h}) \left[ -2 p_1 \cdot p_2
    m_b C^{(5)}_{22} \right.\right.\\
&+& \left.(p_1 \cdot p_3+p_2 \cdot p_3) m_b (2 C^{(5)}_{23}-C^{(5)}_{0})+
    m_b^3 C^{(5)}_{0}-2 m_b m_{\tilde{\chi}_1^{+}}^2 C^{(5)}_{22} - m_b
    \epsilon C^{(5)}_{24} \right]\\
&+& \left. (t,b,\bar{C}^{(3,k)},C^{(3,k)},C^{(5)} \rightarrow
            b,t,\bar{C}^{(4,k)},C^{(4,k)},C^{(6)}) \right\},
\end{eqnarray*}
\begin{eqnarray*}
f_7^{tr,\hat{t}} &=& f_8^{tr,\hat{t}} = f_9^{tr,\hat{t}} = f_{10}^{tr,\hat{t}} = \\
&& \left\{ \frac{-e^2 Q_t^2}{4 \pi^2} \sum_{k=1,2}\left[
    (C^{(3,k)}_{23}-C^{(3,k)}_{22}) (
    V_{H^0\tilde{t}_{k}\tilde{t}_{k}}
    V_{H^0\tilde{\chi}^{+}_{1}\tilde{\chi}^{+}_{1}}^{s} A_{H}+
    V_{h^0\tilde{t}_{k}\tilde{t}_{k}}
    V_{h^0\tilde{\chi}^{+}_{1}\tilde{\chi}^{+}_{1}}^{s} A_{h})\right] \right.\\
&-& \frac{e^2 Q_b^2}{4 \pi^2}
    \left[ m_b (V_{H^0bb} V_{H^0\tilde{\chi}^{+}_{1}\tilde{\chi}^{+}_{1}}^{s} A_{H}+
    V_{h^0bb} V_{h^0\tilde{\chi}^{+}_{1}\tilde{\chi}^{+}_{1}}^{s} A_{h})
    (C^{(5)}_{0}+4 C^{(5)}_{22}-4 C^{(5)}_{23}) \right] \\
&+& \left. (t,b,C^{(3,k)},C^{(5)} \rightarrow b,t,C^{(4,k)},C^{(6)}) \right\},
\end{eqnarray*}
\begin{eqnarray*}
f_{21}^{tr,\hat{t}} &=& f_{22}^{tr,\hat{t}} =
\left\{ \frac{-i m_b e^2 Q_b^2}{4 \pi^2} C^{(5)}_{0} \left[ (V_{A^0bb}
   V_{A^0\tilde{\chi}^{+}_{1}\tilde{\chi}^{+}_{1}}^{ps} A_{A}+V_{G^0bb}
   V_{G^0\tilde{\chi}^{+}_{1}\tilde{\chi}^{+}_{1}}^{ps} A_{G})
   + (t,b,C^{(5)} \rightarrow b,t,C^{(6)}) \right] \right\},
\end{eqnarray*}
$$
f_{i}^{tr,\hat{t}}=0,~~(i=3 \sim 6, 11 \sim 20),
$$
where $\bar{C}^{(3,k)}_{24}=C^{(3,k)}_{24}-\frac{\Delta}{4}$ and
$\bar{C}^{(4,k)}_{24}=C^{(4,k)}_{24}-\frac{\Delta}{4}$.
The form factors in renormalized amplitude of the self-energy corrections
${\cal M}^{s,\hat{t}}$ from Fig.1(f) including t-channel, are expressed as:
\begin{eqnarray*}
f_i^{s,\hat{t}} = 0,~~~~(i=2 \sim 4, 6 \sim 10, 12 \sim 22),
\end{eqnarray*}
\begin{eqnarray*}
f_1^{s,\hat{t}} &=&
\frac{-i e^2 A_t^2}{16 \pi^2} p_1 \cdot p_3 \sum_{k=1,2} (-B^{(3,k)}_{1}
    m_{\tilde{\chi}_1^{+}} F_{1}^{b\tilde{t}_{k}\tilde{\chi}^{+}_{1}}+B^{(3,k)}_{0}
    m_b F_{2}^{b\tilde{t}_{k}\tilde{\chi}^{+}_{1}}\\
&& +B^{(4,k)}_{1} m_{\tilde{\chi}_1^{+}} F_{1}^{t\tilde{b}_{k}\tilde{\chi}^{+}_{1}}-
    B^{(4,k)}_{0} m_t F_{2}^{t\tilde{b}_{k}\tilde{\chi}^{+}_{1}})-i e^2 A_t^2
    p_1 \cdot p_3 \left[ C^{-}_{S} \right.\\
&& +\left. C^{+}_{S}-m_{\tilde{\chi}_1^{+}} (C_{L}+C_{R}) \right]
\end{eqnarray*}
\begin{eqnarray*}
f_{11}^{s,\hat{t}} &=& \frac{f_{5}^{s,\hat{t}}}{2} =
\frac{-i e^2 A_t^2}{16 \pi^2} \sum_{k=1,2} \left[ B^{(3,k)}_{1}
    (m_{\tilde{\chi}_1^{+}}^2-p_1 \cdot p_3)
    F_{1}^{b\tilde{t}_{k}\tilde{\chi}^{+}_{1}} \right. \\
&& -B^{(3,k)}_{0} m_b m_{\tilde{\chi}_1^{+}}
    F_{2}^{b\tilde{t}_{k}\tilde{\chi}^{+}_{1}} \\
&&  \left. -B^{(4,k)}_{1} (m_{\tilde{\chi}_1^{+}}^2-p_1 \cdot p_3)
    F_{1}^{t\tilde{b}_{k}\tilde{\chi}^{+}_{1}}+B^{(4,k)}_{0} m_t
    m_{\tilde{\chi}_1^{+}} F_{2}^{t\tilde{b}_{k}\tilde{\chi}^{+}_{1}} \right]\\
&-& i e^2 A_t^2 \left[ (m_{\tilde{\chi}_1^{+}}^2-p_1 \cdot p_3) (C_{L}+C_{R})-
    m_{\tilde{\chi}_1^{+}} (C^{-}_{S}+C^{+}_{S}) \right]
\end{eqnarray*}
\par
In this work we adopted the definitions of two-, three-, four-point one-loop
Passarino-Veltman integral functions as shown in reference\cite{s19} and
all the vector and tensor integrals can be deduced in the forms of scalar
integrals \cite{s20}.

\begin{center}
{\large \bf Figure Captions}
\end{center}

\parindent=0pt

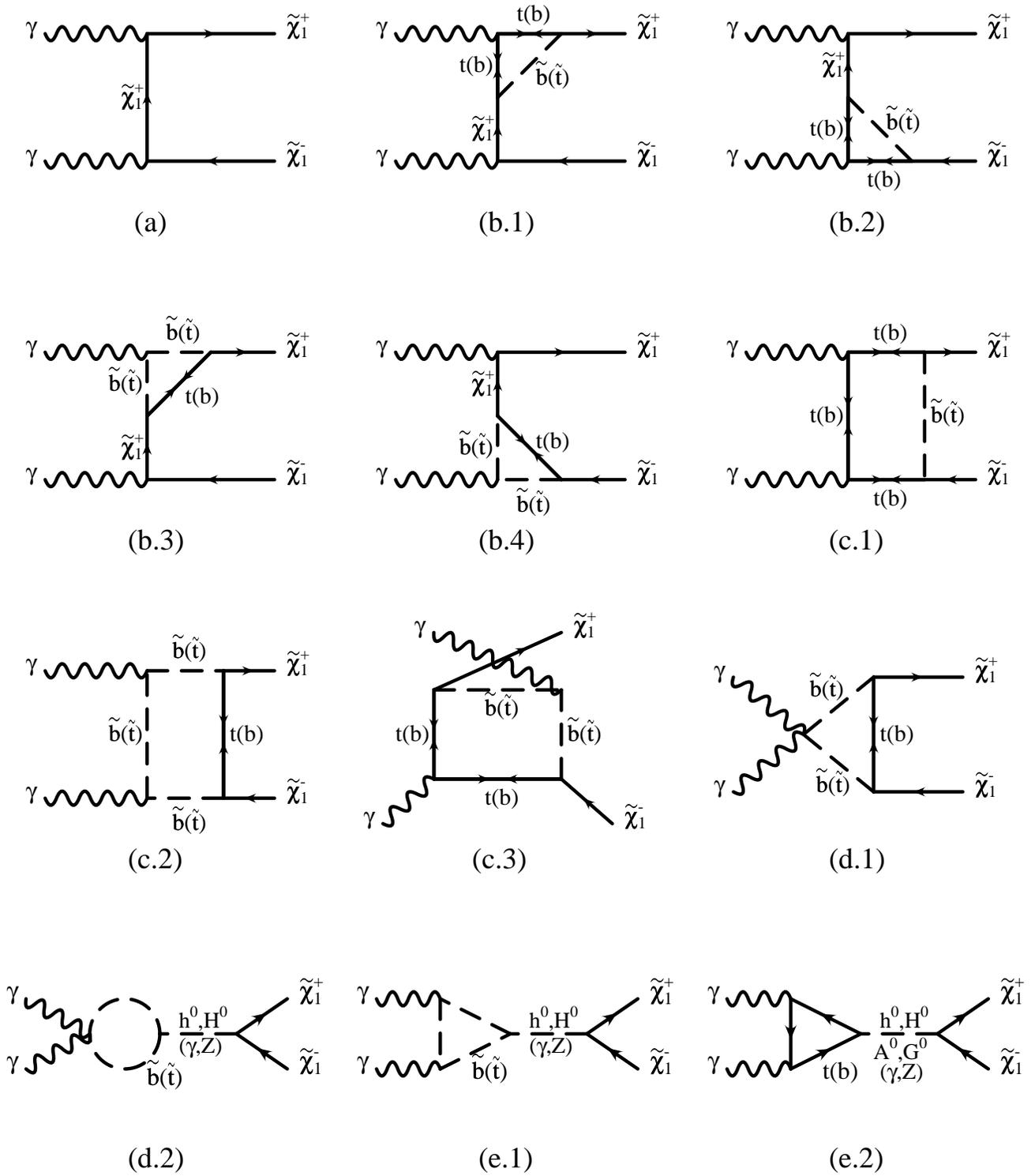
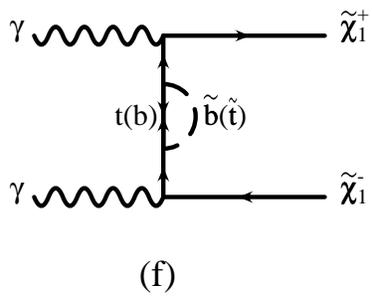
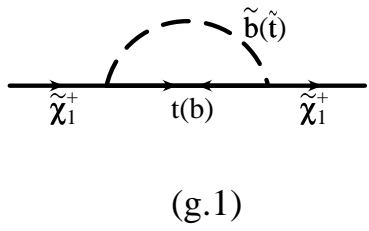
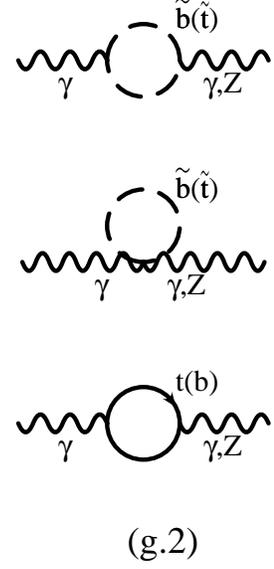
{\bf Fig.1} Feynman diagrams involving one-loop corrections of virtual
     heavy quarks and squarks to the subprocess $\gamma \gamma \rightarrow
     \tilde{\chi}_1^{+} \tilde{\chi}_1^{-}$.
     (a) tree-level diagram.
     (b) $\gamma \tilde{\chi}_1 \tilde{\chi}_1$ vertex corrections.
     (c) box diagrams. (d) quartic interaction corrections. (e) triangle
     diagrams. (f) self-energy diagram. (g) chargino, $\gamma$ and $\gamma-Z^0$
     self-energy diagrams. All the u-channel diagrams are not shown here.

{\bf Fig.2} The Feynman rules for chargino-squark-quark couplings.

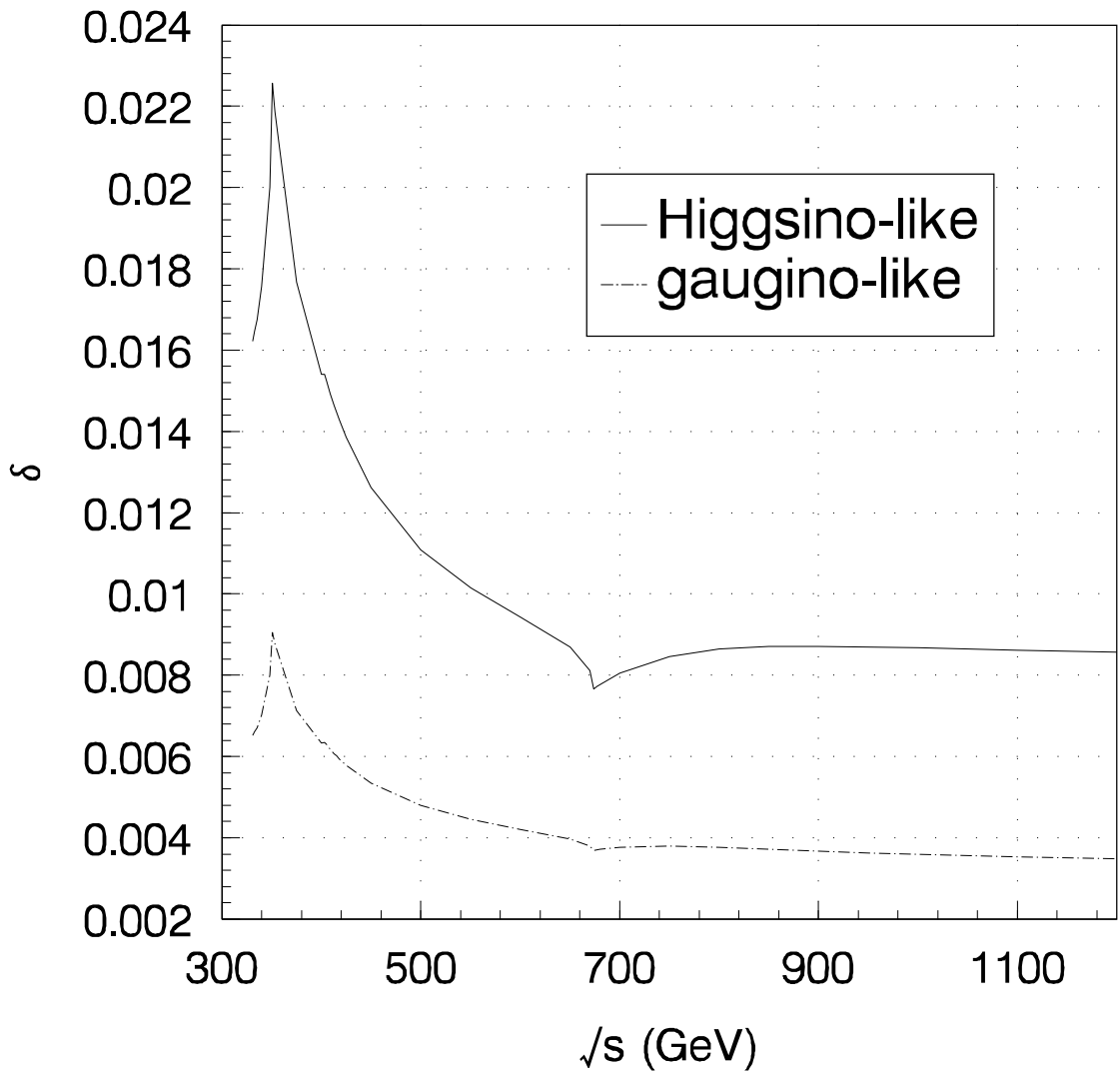
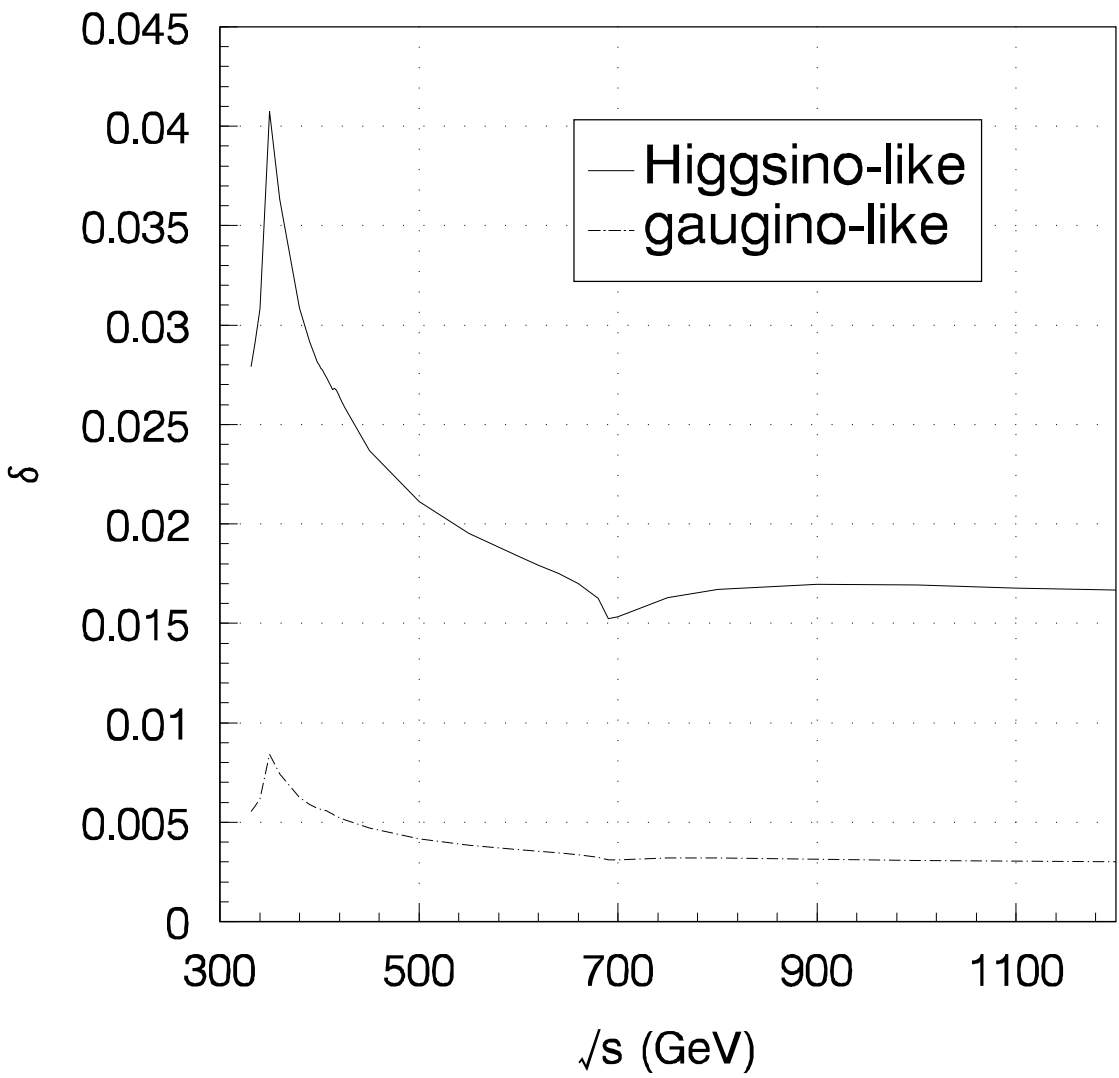
{\bf Fig.3} The relative correction $\hat{\delta}$ of virtual quarks and
     squarks to the subprocess $\gamma\gamma \rightarrow
     \tilde{\chi}_1^{+} \tilde{\chi}_1^{-}$ versus the c.m.s. energy of
     incoming photons $\sqrt{\hat{s}}$ with $m_{\tilde{\chi}^{+}_{1}}=165~GeV$,
     $m_{\tilde{\chi}^{+}_{2}}=750~GeV$, $\tilde{M}=200~GeV$,
     $\phi_{\mu}=\phi_{\tilde{t}}=\phi_{\tilde{b}}=0$ and $m_{A}=150~GeV$.
     (a) $\tan{\beta}=4$. The full-line is for Higgsino-like chargino case.
     The dashed-line for gaugino-like chargino case.
     (b) $\tan{\beta}=40$. The full-line is for Higgsino-like chargino case.
     The dashed-line for gaugino-like chargino case.

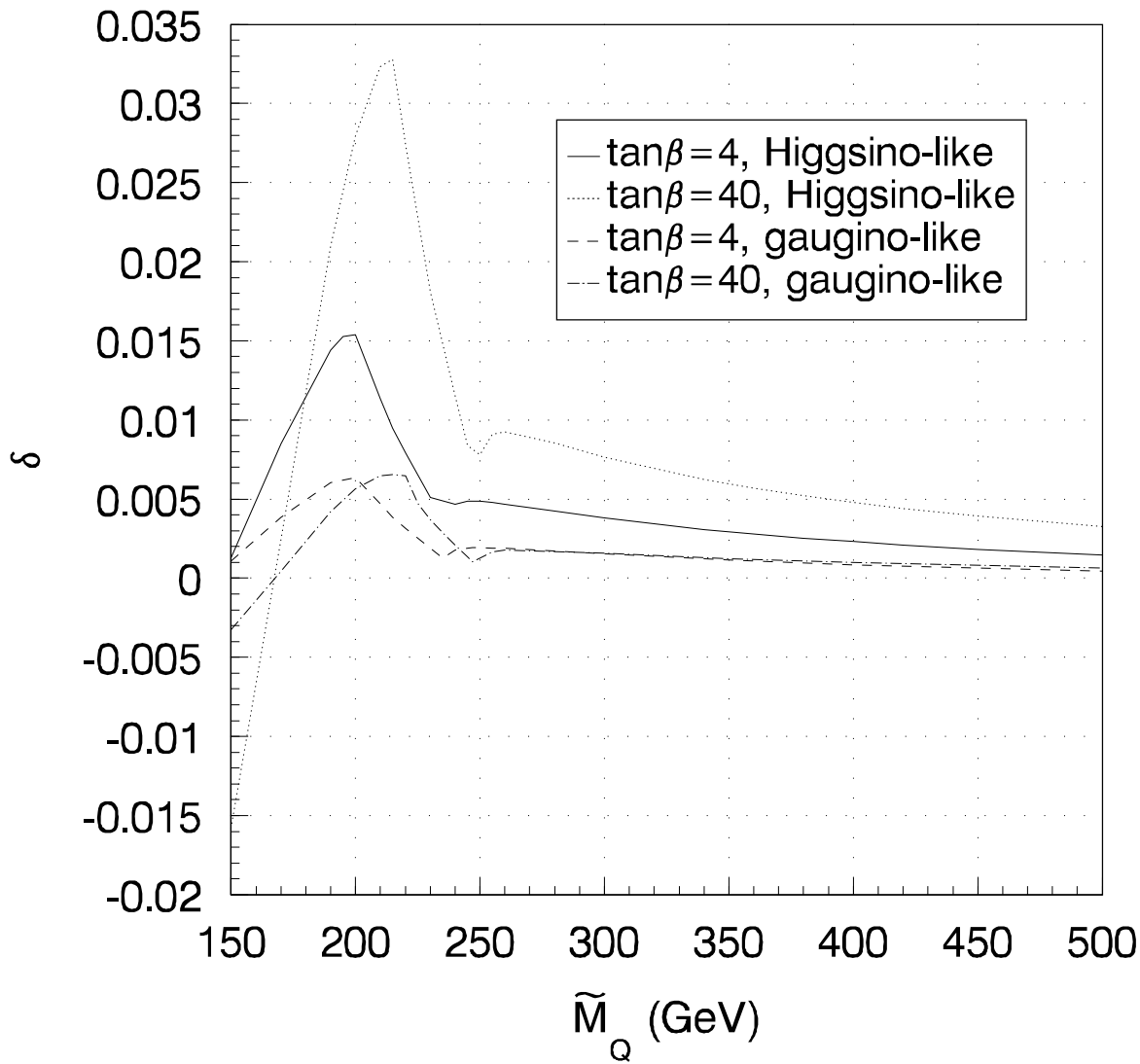
{\bf Fig.4} The relative correction $\hat{\delta}$ of virtual quarks and
     squarks to the subprocess $\gamma\gamma \rightarrow
     \tilde{\chi}_1^{+} \tilde{\chi}_1^{-}$ versus the $\tilde{M}(M_{\tilde{Q}}
     = M_{\tilde{t}}=M_{\tilde{b}}= \tilde{M})$ with $\sqrt{\hat{s}}=400~GeV$,
     $m_{\tilde{\chi}^{+}_{1}}=165~GeV$, $m_{\tilde{\chi}^{+}_{2}}=750~GeV$,
     $\phi_{\mu}=\phi_{\tilde{t}}=\phi_{\tilde{b}}=0$, $\tilde{M}=200~GeV$ and
     $m_{A}=150~GeV$. In Higgsino-like chargino case, the full-line is for
     $\tan{\beta}=4$ and the dotted-line is for $\tan{\beta}=40$.
     $m_{A}=150~GeV$. In gaugino-like chargino case, the dashed-line is for
     $\tan{\beta}=4$ and the dash-dotted-line is for $\tan{\beta}=40$.

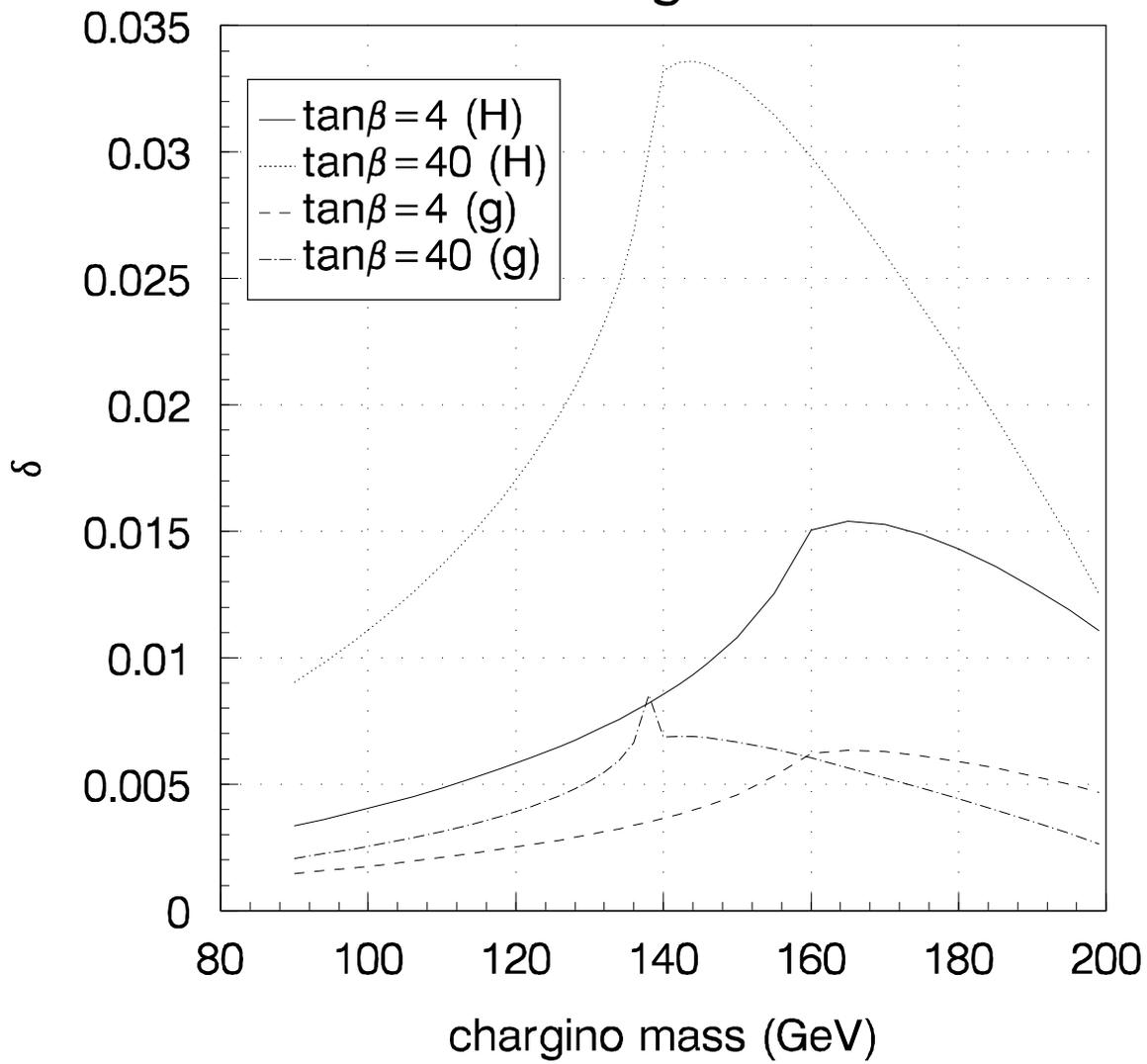
{\bf Fig.5} The relative correction $\hat{\delta}$ of virtual quarks
     and squarks to the subprocess $\gamma\gamma \rightarrow
     \tilde{\chi}_1^{+}\tilde{\chi}_1^{-}$ versus $m_{\tilde{\chi}^{+}_{1}}$
     with $\sqrt{\hat{s}}=400~GeV$, $m_{\tilde{\chi}^{+}_{2}}=750~GeV$,
     $\phi_{\mu}=\phi_{\tilde{t}}=\phi_{\tilde{b}}=0$, $\tilde{M}=200~GeV$
     and $m_{A}=150~GeV$. In Higgsino-like chargino case, the full-line is for
     $\tan{\beta}=4$ and the dotted-line is for $\tan{\beta}=40$.
     $m_{A}=150~GeV$. In gaugino-like chargino case, the dashed-line is for
     $\tan{\beta}=4$ and the dash-dotted-line is for $\tan{\beta}=40$.

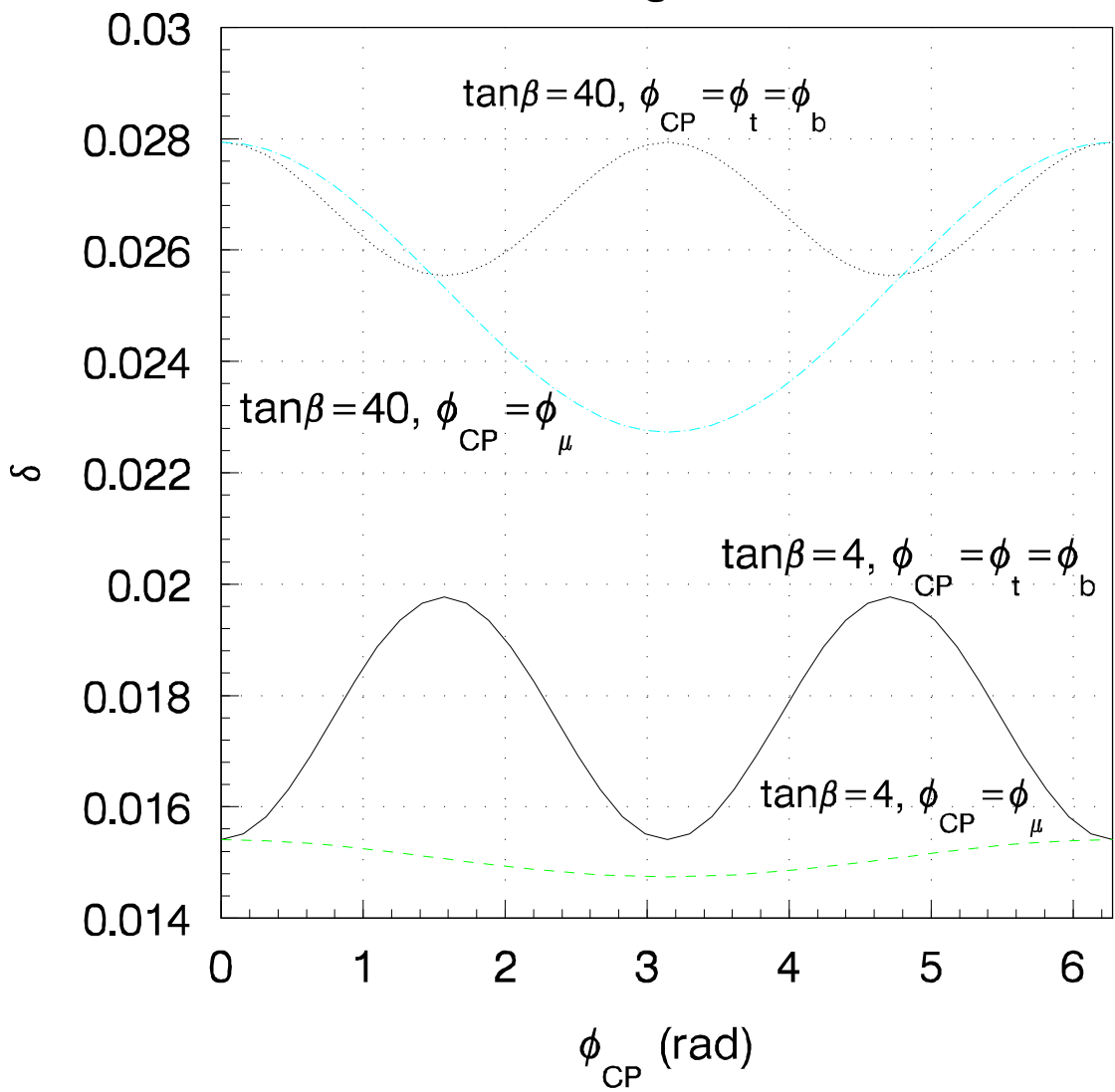
{\bf Fig.6} The relative correction $\hat{\delta}$ of virtual quarks and
     squarks to the Higgsino-like chargino pair production subprocess
     $\gamma\gamma \rightarrow \tilde{\chi}_1^{+}\tilde{\chi}_1^{-}$
     versus $\phi_{CP}$'s with $\sqrt{\hat{s}}=400~GeV$,
     $m_{\tilde{\chi}^{+}_{1}}=165~GeV$,
     $m_{\tilde{\chi}^{+}_{2}}=750~GeV$, $\tilde{M}=200~GeV$ and $m_{A}=150~GeV$.
     The full-line is for $\tan{\beta}=4$ with $\phi_{CP}=\phi_{\tilde{t}}=
     \phi_{\tilde{b}}$. The dashed-line is for $\tan{\beta}=4$ with
     $\phi_{CP}=\phi_{\mu}$. The dotted-line is for $\tan{\beta}=40$
     with $\phi_{CP}=\phi_{\tilde{t}}=\phi_{\tilde{b}}$. The
     dash-dotted-line is for $\tan{\beta}=40$ with $\phi_{CP}=\phi_{\mu}$.

{\bf Fig.7(a)} The cross section $\sigma$ of the process of the Higgsino-like
     lightest chargino pair production via photon-photon fusion in $e^+~e^-$
     collider versus the colliding energy $\sqrt{s}$ in electron-positron
     c.m.s. system with $m_{\tilde{\chi}^{+}_{1}}=165~GeV$,
     $m_{\tilde{\chi}^{+}_{2}}=750~GeV$, $\tilde{M}=200~GeV$,
     $\phi_{\mu}=\phi_{\tilde{t}}=\phi_{\tilde{b}}=0$ and
     $m_{A}=150~GeV$. The full-line is for $\tan{\beta}=4$.
     The dotted-line is for $\tan{\beta}=40$.

{\bf Fig.7(b)} The relative correction of the process $e^+ e^- \rightarrow
     \gamma\gamma \rightarrow \tilde{\chi}_1^{+}\tilde{\chi}_1^{-}$ versus
     the colliding energy $\sqrt{s}$ in electron-positron c.m.s. system with
     $m_{\tilde{\chi}^{+}_{1}}=165~GeV$, $m_{\tilde{\chi}^{+}_{2}}=750~GeV$,
     $\tilde{M}=200~GeV$, $\phi_{\mu}=\phi_{\tilde{t}}=\phi_{\tilde{b}}=0$
     and $m_{A}=150~GeV$. In Higgsino-like chargino case, the full-line is for
     $\tan{\beta}=4$ and the dotted-line is for $\tan{\beta}=40$.
     $m_{A}=150~GeV$. In gaugino-like chargino case, the dash-dotted-line is
     for $\tan{\beta}=4$ and the dash-dotted-line is for $\tan{\beta}=40$.

\end{large}
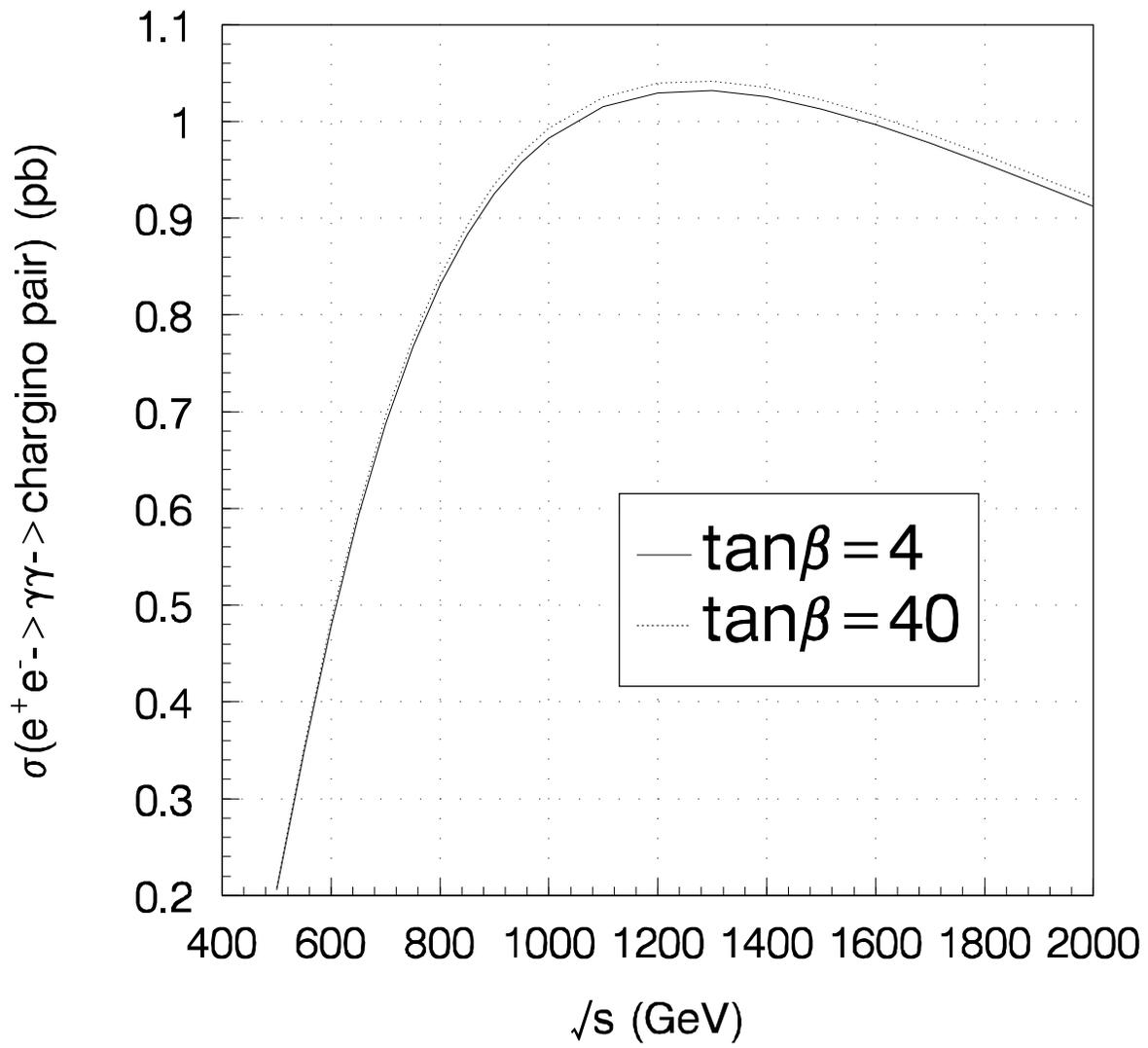
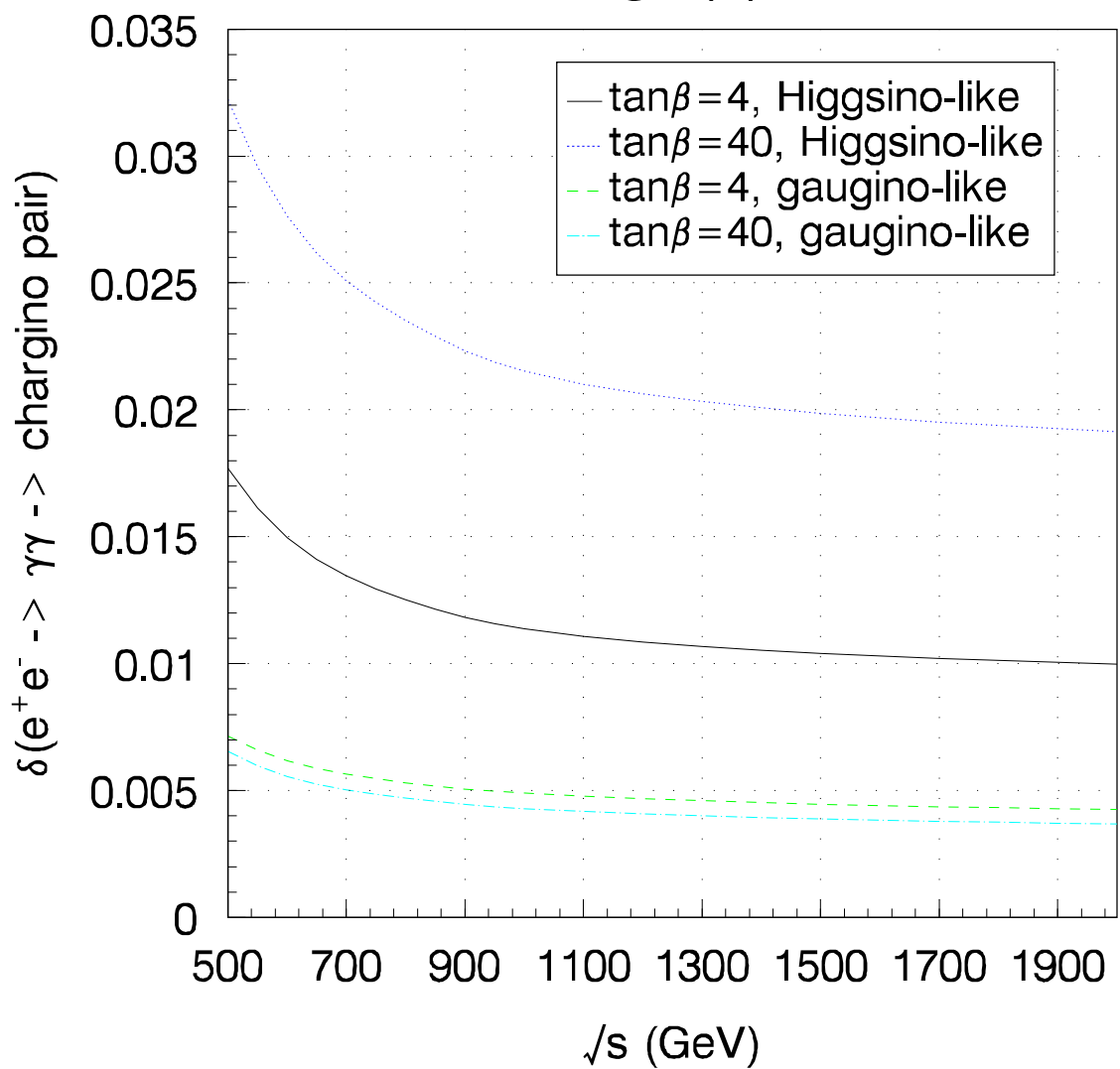
\end{document}